\documentclass[11pt,english]{article}
\usepackage{amssymb,amsmath,graphicx,multicol}
\usepackage{verbatim}
\usepackage[T1]{fontenc}
\usepackage{subfigure}
\usepackage{graphicx}
\usepackage{amssymb}
\usepackage{esint}
\usepackage{a4wide}
\usepackage{psfrag}
\usepackage{babel}
\usepackage{color}
\newcommand{\be}{\begin{equation}} 
\newcommand{\ee}{\end{equation}}
\newcommand{\bea}{\begin{eqnarray}}
\newcommand{\eea}{\end{eqnarray}}
\newcommand{\bi}{\begin{itemize}}
\newcommand{\bra}[1]{\langle #1|}
\newcommand{\ket}[1]{|#1\rangle}
\newcommand{\aaaa}{\mathfrak{a}} 
\newcommand{\ei}{\end{itemize}} 
\newcommand{\bc}{\begin{column}{0.50\textwidth}}
\newcommand{\ec}{\end{column}}
\newcommand{\bcs}{\begin{columns}} 
\newcommand{\ecs}{\end{columns}}
\newcommand{\Tr}[1]{\operatorname{Tr}(#1)}

\newcommand{\Prob}{\operatorname{Prob}}
\newcommand{\meanO}{\mathcal{O}_{aa}}
\newcommand{\iniE}{\overline{E}}
\newcommand{\inie}{\overline{e}}
\newcommand{\Ot}{\mathcal{O}_t}

\makeatletter

\@addtoreset{equation}{section}

\@addtoreset{figure}{section}

\@addtoreset{table}{section}
\makeatother
\begin{document}

\begin{center}
\LARGE
Quench Dynamics in Randomly Generated \\ Extended Quantum Models  
\end{center}
\vspace{0.5cm}
\begin{center}
{\large G. P. Brandino$^{1,2,}$\footnote{Current address: Institute for Theoretical Physics, Universiteit van Amsterdam, Science Park 904, 1090 GL Amsterdam, The Netherlands}, A. De Luca$^{1,2}$, R.M. Konik$^{3}$,  
G.\ Mussardo$^{1,2,4}$
\vspace{0.9cm}}

{\sl $^1$International School for Advanced Studies (SISSA),\\
Via Bonomea 265, 34136, Trieste, Italy\\[2mm]
$^2$INFN, Sezione di Trieste\\[2mm]
$^3$ Condensed Matter and Material Science Department, \\Brookhaven National
Laboratories  Upton, NY USA\\[2mm]
$^4$The Abdus Salam International Centre for Theoretical Physics, Trieste, Italy \\

}

\end{center}

\vspace{0.5cm}
\begin{center}
{\bf Abstract}\\[5mm]
\end{center}
We analyze the thermalization properties and the validity of the Eigenstate Thermalization Hypothesis in a generic class 
of quantum Hamiltonians where the quench parameter explicitly breaks a $Z_2$ symmetry. Natural realizations of such systems are 
given by random matrices expressed in a block form where the terms responsible for the quench dynamics are 
the off-diagonal blocks. Our analysis examines both dense and sparse random matrix realizations of the 
Hamiltonians and the observables. Sparse random matrices may be associated with local quantum Hamiltonians and 
they show a different spread of the observables on the energy eigenstates with respect to the dense ones. In 
particular, the numerical data seems to support the existence of rare states, i.e. states where the observables 
take expectation values which are different compared to the typical ones sampled by the micro-canonical distribution. 
In the case of sparse random matrices we also extract the finite size behavior of two different time scales 
associated with the thermalization process.

\newpage

\section{Introduction}
Largely triggered by recent experiments on cold atoms \cite{Weiss,Hofferberth,Weiler,Greiner,Sadler}, there has been in 
the past few years intense theoretical activity aimed at understanding the non-equilibrium dynamics in closed 
interacting quantum systems following a change in one of the system parameters.
In many interesting cases, the system parameters can be changed rapidly with respect to all other time scales of the system 
that it is meaningful to consider the limit known as a {\em quantum quench}: namely, the system is prepared in an 
energy eigenstate $|\psi_0 \rangle$ of an initial pre-quench Hamiltonian, $H_{pre}$, and then is allowed to evolve 
according to a new post-quench Hamiltonian, $H_{post}$, which differs from $H_{pre}$ by some variation of a parameter. 
Given that such a time evolution is purely unitary, one 
may wonder whether the system reaches, for large times, a new steady state and, moreover, if measurements done on this 
state are able to be related to the thermal density matrix of the system.

Recent progress in understanding thermalization of an extended quantum system
following a quench has involved both analytical and numerical studies. From the
analytical point of view, one of the first results can be traced back to von
Neumann \cite{vonNeumann} (see also \cite{Goldstein}), who pointed out the subtleties involved in defining a quantum
generalization of the notion of ergodicity. Imagine, for example, that with an
appropriate choice of the pre-quench Hamiltonian, $H_{pre}$, we have prepared the
system in a linear superposition of ${\mathcal N}$ energy eigenstates, $|E_a\rangle$, 
of the post-quench Hamiltonian $H_{post}$, all in a shell of energies,
$|E_a-\iniE| < \Delta$, centered at $\iniE$:
\begin{equation}
| \Psi \rangle \,=\,\sum_{|E_a-\iniE| < \Delta} c_{a} | E_a \rangle \,\,\,.
\end{equation}
The time average of the density matrix based on this state, given by
\begin{equation}
\rho_{diag}(\iniE)\,=\,\overline {| \Psi(t) \rangle \langle \Psi(t) |}
\,=\,\sum_{|E_a-\iniE| < \Delta}
| c_a |^2 | E_a \rangle \langle E_a | \,\,\,,
\end{equation}
defines the so-called {\em diagonal ensemble} which is, in general, different from the micro-canonical density matrix defined by
\begin{equation}
\rho_{mc}(\iniE) \,=\, \frac{1}{{\mathcal N}} \sum_{|E_a-\iniE| < \Delta}
 | E_a \rangle \langle E_a | \,\,\,.
 \end{equation}
So, unless it happens that $| c_{a} |^2 = 1/{\mathcal N}$, quantum ergodicity is strictly speaking almost never realized. 
Things may be however different if one looks at the long time evolution (and time average) of the expectation value of 
macroscopic observables, ${\mathcal O}$. For these quantities, 
defining $\langle {\mathcal O} \rangle_{diag} = \rm{Tr} ({\mathcal O}
\rho_{diag}(\iniE))$ and $\langle {\mathcal O} \rangle_{mc} = \rm{Tr} ({\mathcal
O} \rho_{mc}(\iniE))$,
it may be true that the identity 
 \begin{equation}
\overline{ \langle \Psi(t) | {\mathcal O} | \Psi(t)\rangle} \,=\,\langle {\mathcal O} \rangle_{diag}
= \langle {\mathcal O} \rangle_{mc} \,\,\, ,
\label{identityensembles}
\end{equation}
indeed holds. One possibility is that the expectation values ${\mathcal
O}_{aa} =\langle E_a | {\mathcal O} | E_a\rangle$ of the macroscopic observables
do not fluctuate between the Hamiltonian eigenstates which are close in
energy. In this case, in fact, the identity (\ref{identityensembles}) holds for
all those initial states which are sufficiently narrow in energy. This is, in a nutshell, 
the scenario known in the literature as the {\em Eigenstate Thermalization Hypothesis}
(ETH)) which was put forward by Deutsch and Srednicki \cite{deu,sre}, based on previous work by Berry \cite{Berry}, 
and which has been recently advocated by Rigol et al. \cite{Rigol} as the mechanism behind the thermalization
 processes in quantum extended systems.

Recently this hypothesis has been put under intense scrutiny by different groups. 
The main emphasis heretofore has been given to the numerical analysis of specific 
models\footnote{Analytic results for quantum quenches have been obtained only for a restricted class of exactly 
solvable lattice models, such as the XY chain, the Ising model or the XXZ quantum spin chain 
\cite{mccoy,sachdev,Rossini,essler,Caux}. Analytic results have been also obtained for systems nearby the 
critical point \cite{gritsev} or for continuous exactly solvable systems, especially in the regime of 
conformal symmetry \cite{cc,cazalilla,mitra}. However it has been argued that the relaxation phenomena of 
these models, ruled by an infinite number of conserved quantities, may be different from the thermalization 
of a generic model and may require the introduction of a generalized Gibbs ensemble, as proposed in 
\cite{Rigol07} (see also \cite{fiorettomussardo} for a derivation in integrable field theories). 
 In this paper, however, we will not deal with such systems, but rather address these issues in a separate publication.}, 
such as hard-core bosons \cite{Rigol,rigol09}, the Bose-Hubbard model 
\cite{kollath07}, strongly correlated interacting fermions \cite{manmana07}, the Hubbard model \cite{eckstein09}, one dimensional
spin chains \cite{silva}, etc. 
In this paper, instead of analyzing a 
particular system, we take a different approach. 
Namely, our strategy herein is to study the thermalization properties 
of a {\em generic} class of Hamiltonians and a {\em generic} set of observables. The natural language to approach the 
problem from this point of view is obviously provided by random matrices \cite{Mehta}, which in the following will 
parametrize both the Hamiltonians and the observables\footnote{For simplicity we consider hereafter real symmetric matrices.}. 
In particular we have chosen to study the quantum quenches and the relative thermalization in a class of Hamiltonians given by 
\begin{equation}
H(h) = H_0 + h V \,\,\,,
\label{HamiltoniansH}
\end{equation}
where the quench parameter $h$ is meant to explicitly break a $Z_2$ symmetry of the 'unperturbed' Hamiltonian $H_0$. 
Such Hamiltonians, which are arguably among the simplest examples of quantum systems, may model spin chains in the 
presence of an external magnetic field but, as we shall see later, they may also encode the familiar quantum Ising 
chain in a transverse magnetic field.
Given the relative simplicity of this class of Hamiltonians, studying their quench dynamics may be a useful path to 
extract interesting information on generic properties of non-equilibrium systems, thus disregarding, in doing so, 
all additional complications coming from a richer structure of states of a specific model.   

However, even upon adopting the abstract language of random matrices, an important issue governing
thermalization properties and of which to be mindful
is the locality of the interaction. Indeed, the mechanism behind thermalization 
may be different in the case in which the Hamiltonian is local and when it is not. As argued below, local Hamiltonians 
and local observables are associated to {\em sparse} matrices, i.e. matrices with a small proportion of non-zero entries, 
while non-local Hamiltonians and non-local observables are described by {\em dense} matrices. The two kinds of matrices 
have two different densities of states and it is essentially for this reason that one observes a different behavior of 
the eigenstates under a quench of the parameter $h$.

Important features of quantum quench processes in local systems were discussed in a paper by Biroli et al. \cite{Biroli}, 
in particular the role played by rare fluctuations in the thermalization of local observables. These authors considered 
the existence of certain rare eigenstates -- rare compared to the typical ones sampled by the micro-canonical 
distribution -- but which may be responsible, if properly weighted, for the absence of thermalization observed 
in certain systems. As discussed in more detail later, the presence of such states can be detected by studying 
the spread of the expectation values of the observables on the energy eigenstates, in particular by 
the finite size dependence of the distribution of expectation values. The numerical 
analysis that we have performed seems indeed to indicate the existence of these rare states in the case of 
sparse random matrices, while they are absent in the case of dense random matrices. However, in our numerics, 
thermalization is observed nonetheless in sparse matrix Hamiltonians, simply because our averaging procedure on the 
different sampling of observables and Hamiltonians does not place a natural exponentially large weight upon the 
rare states, thus enabling them to break thermalization.

It should be underlined that the existence of rare states in the thermodynamic limit has been debated in the literature 
and in particular in a series of papers by Santos and Rigol \cite{sr1,sr2,sr3}.  They have argued that in a portion of the phase
diagram of an extended t-J model with next-nearest-neighbour interactions, rare states are absent.  We will come back to this
conclusion in our presentation of results.

The paper is organized as follows: in Section \ref{Sec_locality} we discuss the notion of locality in regards to 
a Hamiltonian for a quantum system, and we argue that the corresponding matrix representation is given by a sparse matrix. 
We follow this in Appendix A with an extended discussion of whether a matrix sparse in a position basis is also sparse in the 
corresponding momentum basis.
In Section \ref{ETH} we address the issue of the ETH and, 
following Ref.\,\cite{Biroli}, we briefly remind the reader of certain salient issues regarding the application of 
the ETH to our quenches.
Section \ref{Z2H} is devoted to the discussion of our quench protocol 
implemented on $Z_2$ breaking quantum Hamiltonians. Section \ref{densematrices} deals with this protocol as applied 
to dense random matrices, while Section \ref{sparsematrices} treats
sparse random matrices. In Section \ref{timescales} we discuss the relevant time scales for thermalization 
in sparse random matrices, while our conclusions are presented in Section \ref{Conclusions}.

\section{Locality}
\label{Sec_locality}
In this section we discuss the nature of Hamiltonian matrices associated with {\em local} models. In order to 
consider finite-size matrices, we will always have in mind that the model contains (effectively) proper physical cut-offs 
(for example, volume or lattice spacings) such that the Hilbert space involved is actually finite. For local models 
here we mean models local in the space variable, ${\bf x}$, (if continuous) or in the lattice site, ${\bf i}$, if discrete. 
The main idea of this section is the following: in a generic basis of the Hilbert space, the matrix representation of 
a local Hamiltonian corresponds to a {\em dense} matrix, i.e. a matrix which has all entries different from zero 
(an explicit example will be given below). However, if the theory is local, there will exist a basis 
(in the following called the {\em local basis}), in which the Hamiltonian will be represented by a {\em sparse} matrix, 
i.e. a matrix where the great majority of its entries are zero. In the following, for simplicity, we focus 
our attention on one-dimensional systems. 

\subsection{Local basis} 
To clarify the concept of the local basis, it is worth starting our discussion
by a paradigmatic example: the 1d quantum Ising model in a longitudinal field (generically
a non-integrable model).
In this case the quantum Hamiltonian for $L$ sites is given in terms of Pauli matrices and takes the form:
\begin{equation}
\label{Ising}
 H \equiv \sum_{i=1}^L \sigma^z_i \sigma^z_{i+1} + h \sigma^x_i + \sigma^z_i = \sum_{i=1}^L H_i.
\end{equation}
The last equality makes evident the local nature of this model: the Hamiltonian
has been written as a sum of operators involving only two lattice sites. 
The matrix representation can be obtained once we fix a basis. A typical choice is given by the set of common 
eigenstates of the $\sigma^z_i$ operators.
They can be written as,
\be 
\ket{\uparrow \uparrow \ldots \uparrow}, \quad \ket{\uparrow \uparrow \ldots
\downarrow},  \ldots \quad \Rightarrow \quad \ket{m_1 m_2 \ldots m_L } ,
\label{vectorIsing}
\ee
where each $m_i \in \{\uparrow,\downarrow\}$ corresponds to the two
possible eigenstates of $\sigma^z_i$:
$$ \sigma^z_i \ket\uparrow = \ket\uparrow, \quad \sigma^z_i
\ket\downarrow = - \ket\downarrow \quad\Rightarrow \quad \sigma_z^i
\ket{m_i} = m_i \ket{m_i} .$$
Therefore the Hilbert space is made of $N = 2^L $ elements.
We call this the real-space basis and it can be characterized by the fact that its elements are tensor products of 
the states of each site.
%I removed the footnote -- it seemed too general and I am uncertain at what set of conclusions we
%are claiming at which we have universally arrived indpendent of basis choice.
%\footnote{Clearly this characterization is not unique, but with all the other possible choices sharing 
%this same property, one arrives at the same set of conclusions.} 
Now let's consider the matrix element of the
Hamiltonian density $H_i$:
\be
 H_i^{s,s'} = \bra{m_1 \ldots m_N} H_i \ket {m_1' \ldots m_N'} =
((m_i m_{i+1}+m_i)\delta_{m_i, m_i'} + h\delta_{m_i, -m_i'})\prod_{k\neq i}
\delta_{m_k, m_k'},
\label{nonzeroIsing}
\ee
where $s,s'$ are shorthand for the full set of indices, $m_1,\ldots, m_N$.
From this expression it is easy to deduce that on each row of the matrix there
are $L+1$ non-zero entries and therefore the total number of non-zero elements of
the $N\times N$ matrix $H$  is ${\mathcal N} =  (L+1) N$. Since the total
number of matrix elements is $N^2$, the density of non-zero elements is given
by 
\be 
\rho \,=\,\frac{{\mathcal N} }{N^2} \,=\,\left(\frac{\ln N}{\ln 2}
+1\right)\frac{1}{N} \propto \frac{\ln N}{N}\,\,\,,
\label{densityones}
\ee
while the density of the zeros of the matrix $H$ is 
\be
\rho_0 \,=\,1-\rho\,\simeq\,1-  \frac{k \ln N}{N}.
\label{densityzeros}
\ee
The constant $k$ is related to specific properties of the model, but the
scaling in Eq. (\ref{densityones}) is general. Therefore for large values of $N$, the Hamiltonian matrix $H_N$ 
is a {\em sparse} matrix, i.e. a matrix with a large number of zeros and few non-zero entries. 
This statement holds in general for any local quantum Hamiltonian, as for instance the one coming from a 
quantum field theory: the fraction of non-zero elements of the Hamiltonian is exponentially suppressed 
in the thermodynamic limit $L \rightarrow \infty$. Take for instance a $(1+1)$ bosonic field theory, 
whose Hamiltonian can be written as 
\be 
H \,=\,\int {\mathcal H}(x) \, dx \,=\,\int  \left[\frac{1}{2} \Pi^2(x,t) +
\frac{1}{2} (\partial_x \hat\varphi(x,t))^2
+V(\hat\varphi(x,t)) \right] \,dx \,\,\,,
\label{fieldtheory}
\ee
where $\Pi(x,t) = \frac{\partial\hat\varphi}{\partial t}$ is the canonical
conjugate of the operator $\hat\varphi(x,t)$ entering the equal-time
commutation relation,
\be 
[\hat\varphi(x,t),\Pi(y,t)]\,=\,i \delta(x-y) \,\,\,.
\label{commutationrelation}
\ee
Once the space coordinate $x$ and the field values have been discretized, we can choose the basis of eigenstates 
of the operators $\varphi_{x_i}$ and the Hamiltonian
(\ref{fieldtheory}) can be expressed as a matrix. All the terms except the
momentum become diagonal in this basis. The momentum instead plays the same
role of a "spin flip" operator as $\sigma^x$ does in the Ising case: 
one can easily see that this term provides off-diagonal coefficients in the
matrix, and in each row they will be as many non-zero entries as there are lattice sites. Therefore a
scaling analogous to Eq. (\ref{densityzeros}) is found.

\subsection{The basis of momenta}

In the last section we pointed out the kind of matrices one should expect when a local model is formulated in a
real-space basis. However one can ask the question what happens to the sparseness of the matrix if
the basis is changed.  Among all other possible bases, it is the momentum basis that is often used to express the Hamiltonian
in matrix form. For this reason we analyze here what would be the differences in this case. The details can be found in 
Appendix \ref{appendixMomenta}, where one can see that the sparseness of the Hamiltonian depends on the particularities
of the 
momentum basis that one considers.  But in summary:
\begin{itemize}
 \item If one considers the basis of momenta obtained as a Fourier transform of the rigid
translation of the real-space basis, the Hamiltonian will appear again as a
sparse matrix, perhaps even with a larger density of non-zero entries. This is true
simply because the change of basis we are considering is sparse, i.e. for a state at $L = 4$:
$$ \ket\phi = \ket{\uparrow \uparrow \downarrow \downarrow} \to \ket{\phi_k} = \ket{\uparrow \uparrow \downarrow \downarrow} + e^{\pi i k/2} \ket{\downarrow \uparrow \uparrow \downarrow} +  e^{2\pi i k/2} \ket{\downarrow \downarrow \uparrow \uparrow} +  e^{3\pi i k/2} \ket{\uparrow \downarrow \downarrow \uparrow}$$
and the number of terms in the last expression is always less than $L \ll 2^L$.
 \item If one considers the basis of momenta obtained taking the Fourier transform
of the free single particle excitations:
$$ \aaaa^\dag_k = \sum_{j = 0}^{L-1} e^{i k j} a_j^\dag $$
the change of basis in each block of defined
total momentum will not be sparse. In the general case, the
Hamiltonian matrix will be characterized by dense blocks of fixed total
momentum. 
\end{itemize}
These considerations allow us to stress one point: there does exist a basis, regardless of interactions, 
in which momentum is a good quantum number and which maintains the sparseness of the Hamiltonian matrix.  

After these general remarks, we now focus on the real-space basis keeping the scaling Eq. (\ref{densityzeros}) describing the 
sparseness of the matrix.

\section{Quantum quenches, thermalization, and the ETH}
\label{ETH}
Let us consider an initial state $|\psi_0\rangle$ which is an eigenstate of an initial
Hamiltonian, $H(h^<)$, governed by the parameter $h^<$.  At $t=0$ we abruptly (i.e. non-adiabatically)
change the value of the parameter to $h^>$. The evolution of the initial state will be then governed 
by the dynamics given by $H(h^>)$. Our interest is in the long time behaviour of expectation values of 
some one-point observable, $\langle \psi_0(t)|{\mathcal O}| \psi_0(t)\rangle$. An observable has a thermal 
behavior if its long time expectation values
coincides with the micro-canonical prediction, i.e. 
\be
\langle \psi_0(t) |{\mathcal O} |\psi_0(t)\rangle \xrightarrow{t \to \infty} \Tr{{\mathcal O}
\rho_\text{mc}}\,=\,\langle {\mathcal O} \rangle_\text{mc} \,\,\,.
\label{ltq}
\ee
However as the systems we will consider are finite sized, one has to allow for quantum revivals. 
A proper way to understand the above limit is then to require that eq.\,(\ref{ltq}) holds in the long time limit 
at almost all times. Mathematically, this means that the mean square difference between the LHS and RHS of eq.\,(\ref{ltq})
averaged over long times is vanishingly small for large systems -- a restatement of von Neumann's 
quantum ergodic theorem \cite{vonNeumann,Goldstein}.  We will come back to the time scales and their subtleties
involved in the approach to equilibrium in Section \ref{timescales}.  For now we say that an observable thermalizes if 
its long-time average coincides with the micro-canonical average, namely  
\be
\overline{\langle \psi_0(t) | {\mathcal O} |\psi_0(t)\rangle}
\equiv \lim_{T\rightarrow\infty}\frac{1}{T}\int^T_0 \langle \psi_0(t) | {\mathcal O} |\psi_0(t)\rangle = \sum_a |c_a|^2
{\cal O }_{\alpha\alpha} =  \langle {\mathcal O} \rangle_\text{mc} \,\,\,,
\label{ltq2}
\ee	
where $c_a=\langle \psi_0|E_a\rangle$ are the overlap of the initial state on
the eigenstate $\ket{E_a}$ of $H(h^>)$, and 
$\meanO=\langle E_a|{\mathcal O}|E_a\rangle$ are the expectation values of the
observable, $O$, on the post-quench eigenstates. Eq. (\ref{ltq2}) defines the diagonal ensemble prediction, with the corresponding density matrix defined as 
\be
\rho_\text{diag}= \overline{|\psi_0(t)\rangle \langle\psi_0(t)|}=\sum_a |c_a
|^2|E_a\rangle \langle E_a| \,\,\,,
\ee
supposing the eigenstates of $H(h^>)$ are non-degenerate.

%\subsection{Eigenstate Thermalization Hypothesis}

A possible mechanism for the thermal behavior of physical observables 
is based on the so called {\it Eigenstate Thermalization Hypothesis} (ETH)
\cite{deu,sre}. It states that the expectation value of a physical observable,
$\meanO =\langle E_a |{\mathcal O}|E_a\rangle$, on an
eigenstate, $|E_a\rangle$, of the Hamiltonian is a smooth function of its energy, $E_\alpha$, with its value 
essentially constant on each micro-canonical energy shell. In such a scenario, thermalization in the 
asymptotic limit follows for every initial condition sufficiently narrow in energy.
ETH implies that thermalization can occur in a closed quantum system, different from the classical
case where thermalization occurs through the interactions with a bath.
As pointed out by Biroli et al. \cite{Biroli} there are two possible interpretation of ETH: a weak one, which can be shown 
to be verified even for integrable models, which states that the fraction of non-thermal states vanishes
in the thermodynamic limit, and a strong one which states that non-thermal states completely disappear 
in the thermodynamic limit.  In the weak version of the ETH, not every initial condition will thermalize.

We briefly remind the reader of the origin of these two interpretations as it will be salient
later. Firstly, for thermalization to occur one 
needs a distribution of the overlaps peaked around the energy $E=\langle\psi_0|H|\psi_0\rangle$. 
As shown in Ref. \cite{Rigol}, 
the energy density $e$ has vanishing fluctuations in the thermodynamic limit
\be
\Delta e =\frac{\sqrt{\langle E^2\rangle_\text{diag}-\langle
E\rangle_\text{diag}^2}}{L} \propto \frac{1}{L^{1-\sigma/2}} \to ~0 \text{~for~} L\to\infty
\label{enfluc}
\ee
where $L$ is the system size and $\sigma$ is the dimension of the space over which the
coupling $h$ is adjusted in the quench.  In our random quench we are adjusting the parameter h
over the entire breadth of the system and this system is meant to be a proxy for
a generic one dimensional quench.  Thus $\sigma=1$.
However we will see that $\sigma$ is effectively larger when we consider quenches in dense matrices,
and consequently $\Delta e$ does not vanish in the thermodynamic limit.
Correspondingly this property means that the distribution of
intensive eigenenergies (eigenenergies scaled by $1/L$) with weights $|c_a|^2$ 
is peaked for large system
sizes. If the ETH is true, an immediate consequence of property Eq. (\ref{enfluc}) would
be that averages in the diagonal ensemble coincide with averages in the
micro-canonical ensemble. However, for a finite system, there will always be
finite fluctuations of $\meanO$. To characterize the ETH mechanism we need then
to have some control on the evolution of the distribution of $\meanO$ in 
approaching the thermodynamic limit.  As shown in \cite{Biroli}
the width of the distribution $\meanO$ of an
intensive local observable\footnote{${\mathcal O}$ is an intensive local
observables if it can be written as
$\frac{1}{L}\sum_\alpha {\mathcal O}_\alpha$ where ${\mathcal O}_\alpha$ are finite ranged
observables and the sum is over a local spatial region.} vanishes in the thermodynamic limit
\be
(\Delta {\mathcal O}_e)^2\,=\,\frac{\sum_e \meanO^2}{N_e} - \left(\frac{\sum_e
\meanO}{N_e}\right)^2  \to 0 \text{ for } L \to \infty
\label{Ofluc}
\ee
where $e$ is the intensive energy defining a micro-canonical shell including
$|E_a\rangle$ such that $E_a/L\in [e-\epsilon,e+\epsilon]$ and
$N_e$ is the number of states in the microcanonical shell.
Eq.\, (\ref{Ofluc}) implies that the fraction of states characterized by a value of
 $\meanO$ different from the micro-canonical average vanishes in the
thermodynamic limit. However states with different values 
of $\meanO$ may exist. These states live in the tails of the shrinking ${\mathcal
O}_{aa}$ distribution and are expected to be small in number. This is why they
are called "rare".
These states, however, under proper conditions, can be relevant to the issue of thermalization. 
Indeed, if in the $|c_a|^2$ distribution they are weighted heavily,
the diagonal ensemble average will be different from the micro-canonical and
the system keeps a memory of the initial state. As emphasized in
\cite{Biroli}, it is clear that the weak interpretation of ETH does not imply
thermalization in the thermodynamic limit for every initial condition, while,
with the proviso that Eq. 3.4 holds, the strong interpretation does.

\section{ $\mathbb{Z}_2$ symmetry breaking quench protocol}
\label{Z2H}

The class of Hamiltonians we choose to study can be thought as akin to the quantum Ising
model in the presence of an additional longitudinal field.  The quench protocol involves, for the sake
of specificity, taking $h$ to $-h$.
This quench reflects that the Ising Hamiltonian is not invariant under the $Z_2$ operator, ${\cal P}=e^{i\pi( L/2 + S_z)}$:
$$
{\cal P}H(h){\cal P}^\dagger = H(-h).
$$
In order to mimic this in the context of random matrices, we suppose we have divided the canonical basis of 
Ising states into two groupings, even and odd under ${\cal P}$, and to then have sorted them by ordering all even states 
before any odd states.  In Ising, the transverse field term couples states with different parity, such that the 
dependence of the Hamiltonian on the external field is seen in the off-diagonal blocks, i.e.
\begin{equation}
\label{hamiltonianForm}
H(h)=\left(\begin{array}{cc} A &h B\\ h B^T & C \end{array}\right) \,\,\,. 
\end{equation}
It is this form then that we take for our random matrices.

Observables of the systems associated to the Hamiltonian (\ref{hamiltonianForm}) can be split into even and 
odd $\mathbb{Z}_2$ classes. This classification is again motivated by the case of the Ising-spin chain 
in a transverse magnetic field, where the natural observables $\sigma_z$ and $\sigma_x$ are respectively odd and 
even w.r.t. to the action of ${\cal P}$. The even observables have non-zero elements in the diagonal blocks alone, 
while the odd observables are non-zero only in the off-diagonal blocks:
\be
\label{evenodd}
 E = \frac 1 {L} \left(\begin{array}{cc} A & 0\\ 0& C \end{array} \right), \qquad 
O = \frac 1 {L} \left(\begin{array}{cc} 0 & B\\ B^T& 0 \end{array} \right) \,\,\,, 
\ee
where the volume factor $L$ has been added to make these quantities intensive. Since we expect the Hilbert space to be exponentially large in the volume of the system we fix the system size corresponding to an $N\times N$ random matrix via
$$ L = \ln N \,\,\,.$$
After defining the Hamiltonian as above, we will analyze the quench dynamics under the quench $h\rightarrow -h$.
We will study the long-time behavior of both odd and even observables. In our numerical analysis, we have examined 
different values of the initial and final value of the parameter $h$, and found that in the limits 
$h \ll 1$ and $h \gg 1$, the quench dynamics are essentially trivial because the initial and final Hamiltonian 
share the same eigenvectors.  For this reason, we will discuss only the intermediate case
\be
\label{quenchProtocol}
 h^{\rm pre-quench} = -1 \quad \rightarrow \quad h^{\rm post-quench} = 1 \,\,\,. 
\ee
Given these constraints we will still however consider two cases:
\begin{itemize}
\item In one case we will look at ensembles of sparse random matrices, motivated by the previous 
considerations about the relationship between locality and sparseness. It should be stressed however that 
while a local observable will necessarily be sparse the converse is not necessarily true.
Nevertheless the study of sparse random matrices may provide some reliable insights into some of the 
questions of the thermalization in local Hamiltonians.
\item In the second case we will look at matrices which are dense.  Here then we are deviating from the motivation
provided by the quantum Ising model.
\end{itemize}
In both cases the entries of the random matrices will be generated according to the 
Gaussian orthogonal ensemble (GOE).  

We first consider quenches involving dense matrices.

\section{Thermalization in Dense Random Matrices Ensemble}
\label{densematrices}
%\subsection{Definition of the ensemble}
To define the Hamiltonian in the dense case, we generate three $N/2 \times N/2$ matrices $A,B,C$ 
and then assemble them according to Eq. (\ref{hamiltonianForm}). The matrices $A,C$ are symmetric and 
chosen according to the measure, $\mu(M)$, of a properly normalized GOE ensemble:
$$
\mu(M) \equiv \exp\left(-\frac{N \Tr{M^2}}{4 L^2} \right)\,\,\,, 
$$
while the matrix $B$ has all of its entries distributed according to a normal distribution 
with 0 mean and variance equal to 
$\frac{2L^2}{N}$. 
For  $h = \pm 1$  the Hamiltonian itself will also be distributed according to the GOE ensemble and 
therefore the eigenvalues obey the semicircle law:
\be
\label{semicircle}
\rho(E)=\frac{1}{2\pi L}\sqrt{4L^2-E^2} \,\,\,.
\ee
The spectrum thus falls in the range $[-2L,2L]$ and is therefore extensive as required. 
The observables are obtained with an analogous procedure, generating new matrices $A,B,C$ 
and then using the expressions Eq. (\ref{evenodd}).
The numerical results reported below are calculated according to the following procedure: five instances of the Hamiltonian are generated according to the prescriptions above  and for each instance of the Hamiltonian forty instances of the observable are generated. The relevant quantities are calculated for each instance of the observables, then the results are averaged.

\subsection{Numerical results}
One of the prerequisites for the ETH to operate is that given the initial state $|\psi_0\rangle$ with energy 
$$
\inie \equiv \frac 1 L \langle\psi_0|H_\text{post}|\psi_0\rangle \,, 
$$ 
the structure of its overlaps $|c_a|^2=|\langle\psi_0|E_a\rangle|^2$ with the post-quench eigenstates, 
as a function of the intensive energy $e_a = E_a/L$, is peaked around $\inie$. 

\begin{figure}[t]
\centering
$\begin{array}{cc}
\includegraphics[width=0.4\textwidth]{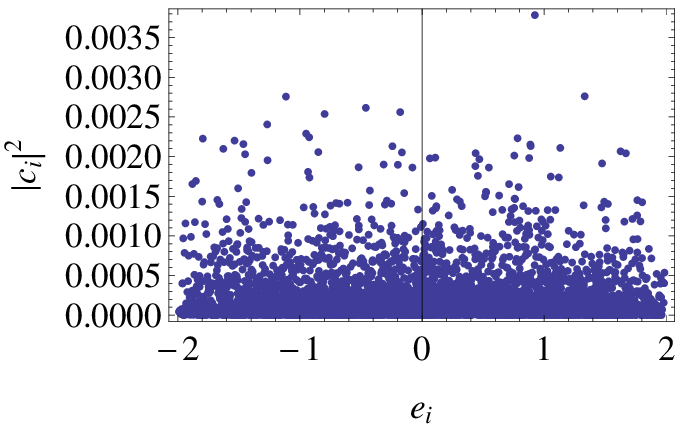} &
\includegraphics[width=0.4\textwidth]{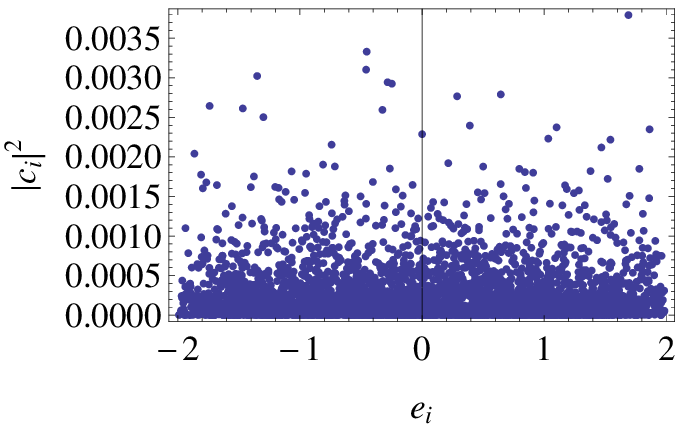}\\
\end{array}$
\caption{Dense random matrices (with  $N=4000$). Overlaps for the quench process. Left: 
$|c_a|^2$ for the 'ground' state.  Right: $|c_a|^2$ 
for the 2000$^\text{th}$ state, in the middle of the energy band.}
\label{sampleoverdense}
\end{figure}

\begin{figure}[b]
\centering
\includegraphics[width=0.4\textwidth]{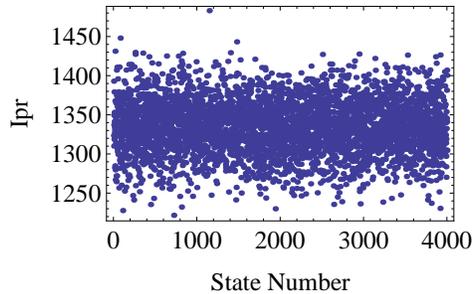} 
\caption{Dense random matrices (with $N=4000$). Typical IPR of the initial states. } 
\label{sampleiprdense}
\end{figure}

\begin{figure}[t]
\centering
$\begin{array}{cc}
\includegraphics[width=0.4\textwidth]{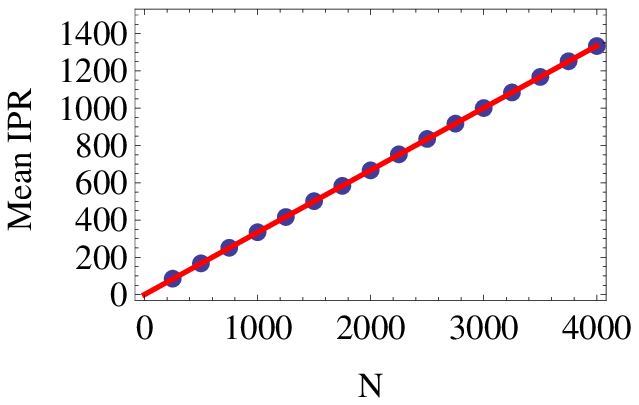} &
\includegraphics[width=0.4\textwidth]{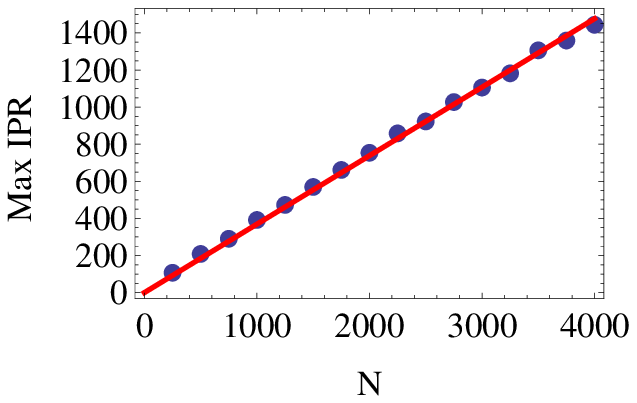}\\
\end{array}$
\caption{Dense random matrices. IPR vs. matrix size. The red lines are linear fits $y=ax$. 
Left panel: average IPR on all initial states ($a=0.3336$). Right panel: maximum IPR ($a=0.3764$). }
\label{iprscalingdense}
\end{figure}

We find this to be not the case for dense matrices, as can be explicitly seen by the two sample states 
in Fig. \ref{sampleoverdense} drawn from the bottom and middle of the
spectrum.
Moreover, calculating the standard deviation of the energy on the initial state, we find that 
it is always large (around 1/4 of the range of the total spectrum) for all initial states, showing that 
the relation Eq. (\ref{enfluc}) does not hold, i.e. the effective dimension $\sigma$ satisfies $\sigma>2$.
The broad distribution of overlaps is confirmed by the analysis of the Inverse Participation Ratio (IPR), defined as 
\be
\text{IPR}\,=\,\frac{1}{\sum_a c_a^4 } \,\,\,.
\ee
We show in Fig. \ref{sampleiprdense} the IPR for the eigenstates of a single realization of a dense matrix.
The IPR in this case is sharply distributed around $N/3$. This finding can be understood
through a simple model of random vectors on a N-sphere of unit radius (Porter-Thomas distribution). By a simple integration one finds \cite{Haake}
\be
\langle c_a^4\rangle \,=\, 3/N^2 \,\,\,,
\ee  
and therefore the IPR scales as 
\be
\frac{1}{\sum_a c_a^4 } \simeq N/3 \,\,\,. 
\ee
This scaling is confirmed by our data, as shown in Fig. \ref{iprscalingdense}.
Moreover, the fact that the mean IPR and the maximum IPR almost coincide is 
confirmation that all initial states are equivalent. This means that the pre-quench and the post-quench bases 
of the energy eigenvectors are completely random with respect one another.  Eigenstates therefore 
have no reason to be localized in energy. 

\begin{figure}[b]
\centering
$\begin{array}{cc}
\includegraphics[width=0.4\textwidth]{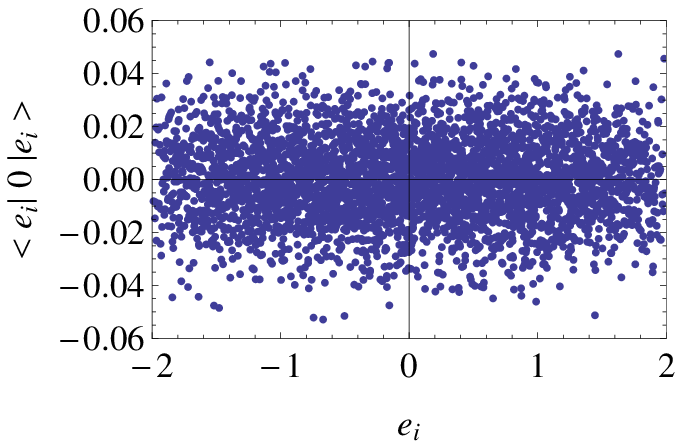} &
\includegraphics[width=0.4\textwidth]{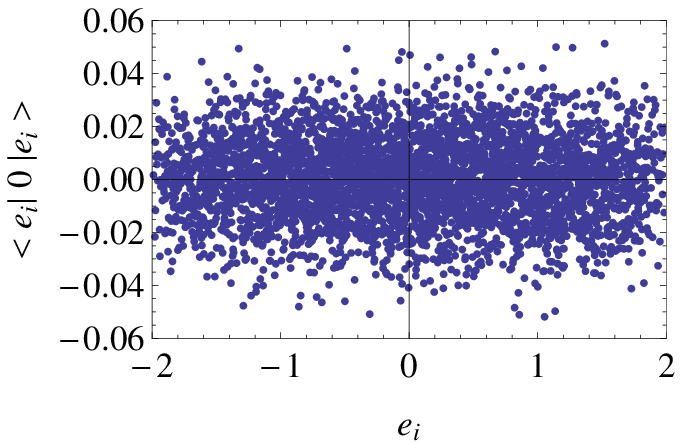}\\
\end{array}$
\caption{Dense random matrices. EEV $\langle E_a|{\mathcal O}|E_a\rangle$ vs
$E_a$. Left: even observable. Right: odd observable.}
\label{EEVsampledense}
\end{figure}

Let us now turn our attention to the expectation values of observables since the main content of the ETH 
concerns the distribution of the eigenstate expectation values (EEVs), $\meanO$, and their behavior 
when the system size is increased. We first report two sample EEV distributions, given in Fig. \ref{EEVsampledense}, 
which show no energy dependence.
We can argue (and we have checked numerically) that the distribution of an intensive observable over the whole energy 
spectrum shrinks to zero for increasing system size. 
Moreover, it is not only the variance but even the support of the distribution of the observables that goes to zero
inasmuch as the difference of the EEV maximum and minimum is going to zero as $N \rightarrow \infty$ (see
Fig. \ref{semicircleintensive}).

More precisely, we have:
\be
\label{observableexpansion}
\meanO \equiv \langle E_a | {\mathcal O} | E_a \rangle = \sum_{\beta}
A_{a,\beta} {\mathcal O}_\beta \,\,\,, 
\ee
where $\beta$ indexes the eigenstates, $\ket{\beta}$, of the observable
${\mathcal O}$, while ${\mathcal O}_\beta$ is the corresponding eigenvalue, and $A_{a,\beta} = |\langle E_a |
\beta \rangle|^2$. To estimate the r.h.s. we argue for an equivalency of observables and hold that
the IPR of an eigenvector $|E_a\rangle$ of the post-quench Hamiltonian relative to the basis
of eigenvectors $|\beta\rangle$ equals the IPR of the initial state $|\psi_0\rangle$ in the basis $|E_a\rangle$.
So we can suppose that $\meanO$ can be expanded in terms of a set of $\frac{N}{3}$
states, each of which is given by
\begin{figure}[t]
\centering
$\begin{array}{cc}
\includegraphics[width=0.4\textwidth]{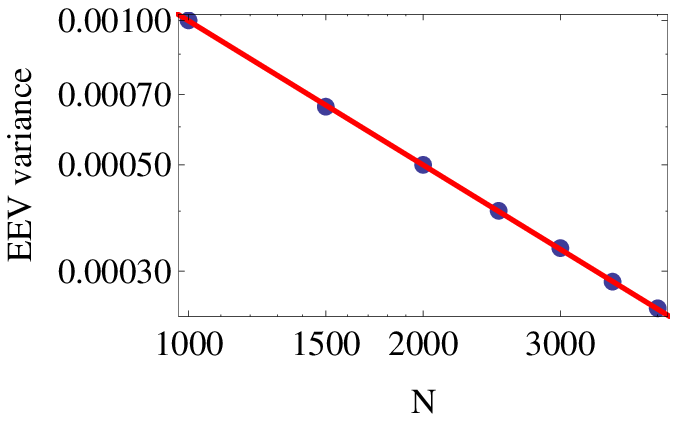} &
\includegraphics[width=0.4\textwidth]{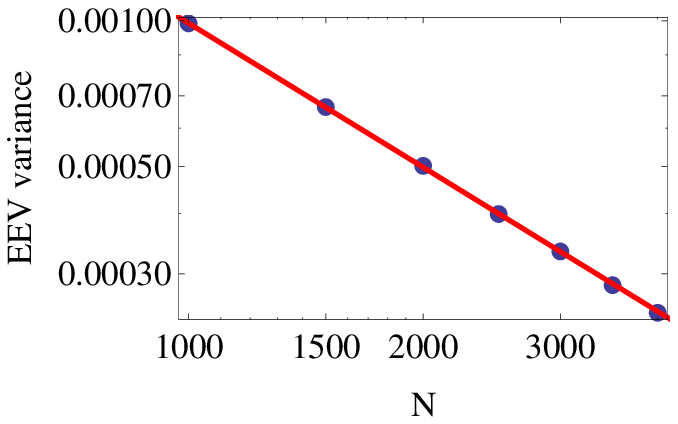}\\
\end{array}$
\caption{Dense random matrices. EEV variance (averaged over the entire spectrum) vs. matrix size. The
continuous line is the fit $a/x^b$ with $ b = 0.9 \pm 0.1$. The data points for the even and
odd observables exactly overlap in this plot.}
\label{EEVvariancedense}
\end{figure}
\be
\label{overlapIPR}
 A_{a, \beta} \simeq \frac{1}{IPR(H \to {\mathcal O})} \simeq \frac{3}{N} \,\,\,. 
\ee
Then, if we assume $A_{a\beta}$ and $A_{a\beta'}$ are independent and note that $\meanO$ has zero mean, we obtain
$$
 \overline{|\meanO|^2} \simeq \frac{N}{3} \left(\frac{3 \sigma_{\mathcal O}}{N} \right)^2 
 \simeq \frac{3 \sigma_{\mathcal O}^2}{N} ,
$$
where $\sigma_{\mathcal O}^2$ is the variance of the spectrum of the observable 
${\mathcal O}$.  In Fig. \ref{EEVvariancedense}, the numerical results are
plotted together with a power-law fit and, 
as expected, the exponent is indeed close to one.

\begin{figure}[b]
\centering
$\begin{array}{cc}
\includegraphics[width=0.4\textwidth]{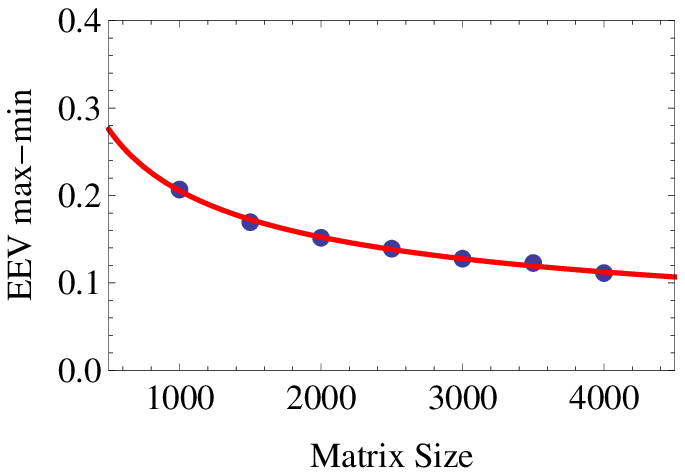} &
\includegraphics[width=0.4\textwidth]{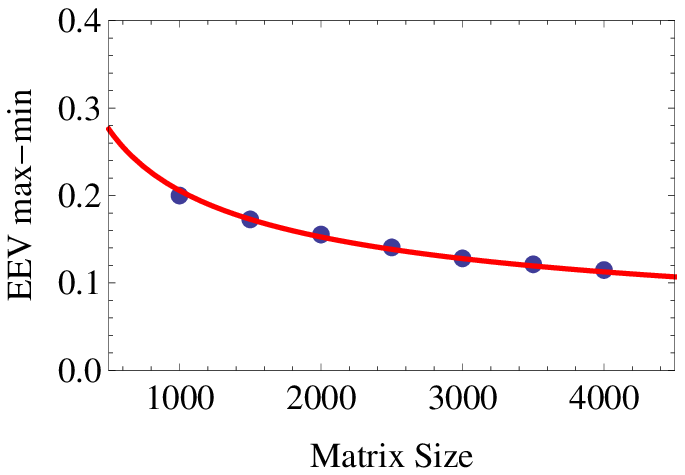}\\
\end{array}$
\caption{Dense random matrices. Max-min EEV vs. matrix size for the range of even (left) and odd (right) observables (again
the behavior of the two is indistinguishable).  
The continuous line is the fit $a\sqrt{\frac{\ln{N}}{N}}$.}
\label{minmaxdense}
\end{figure}

Now let us consider the full support of the distribution of the EEVs where we define
$\delta_{\mathcal O}$ as the difference of the maximum and the minimum of the EEVs 
among all the energy eigenstates $|E_a\rangle$. Since the distribution is symmetric about zero, we have:
\be
\label{maxminspread}
\delta_{\mathcal O} \,=\, 2\max_a \{ \meanO \} \,. 
\ee
To estimate the scaling of this quantity, we again approximate all the overlaps $A_{a,\beta}$ 
as in Eq. (\ref{overlapIPR}). Therefore we are led to estimate the maximum of the quantity
$$ 
{\cal O}_{aa} \equiv \frac{3}{N} \sum_\beta' {\mathcal O}_\beta ,
$$
where the prime on the sum indicates that only $1/3$ of the total $\beta$'s are being summed over.
\begin{figure}[t]
\centering
$\begin{array}{cc}
\includegraphics[width=0.4\textwidth]{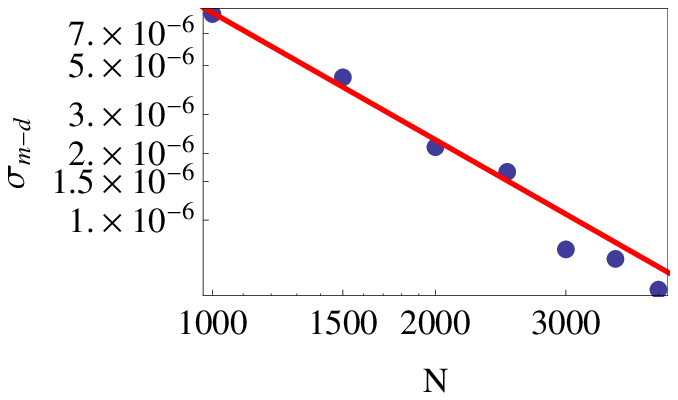} &
\includegraphics[width=0.4\textwidth]{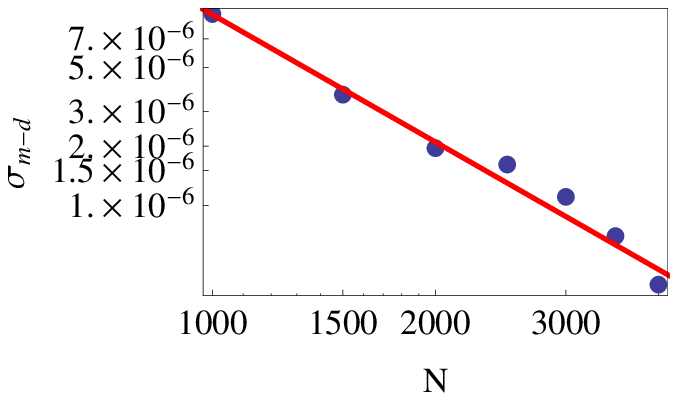}\\
\end{array}$
\caption{Dense random matrices. $\sigma=\sqrt{({\mathcal O}_\text{micro} -
{\mathcal O}_\text{diag})^2}$ vs. matrix size for initial states laying in the central part of the 
spectrum $\overline{e}\approx 0$. The continuous line is the fit $a/x^b$. Left: even observable $b=1.9\pm 0.1$; 
right: odd observable $b=2.1\pm 0.1$}
\label{centerscalingdense}
\end{figure}
As before we suppose that the random variables ${\mathcal O}_\beta$ are independently distributed 
according to the intensive semicircle law (${\mathcal O}_\beta \in [-2,2]$):
\be
\label{semicircleintensive}
\rho(x) \equiv \operatorname{Prob}({\mathcal O}_\beta = x) \,=\, 
\frac{1}{2\pi} \sqrt{4 - x^2} \,\,\,. 
\ee
In this case we can use large deviation theory (see for instance \cite{largedeviation}) and it follows that
\be
\label{largedeviation}
 \Prob({\cal O}_{aa} > x) \simeq e^{- N I(x)} \,\,\,, 
\ee
where $I(x)$ is the \textit{rate function} given by
\be
\label{Idef}
 I(x) = \sup_\theta [\theta x - \lambda(\theta)] \,\,\,. 
\ee
The function $\lambda(x)$ is the cumulant generating function 
\be
\label{lambdadef}
 e^{\lambda(\theta)} \,= \, \int_{-2}^2 \rho(x) e^{x \theta} = \frac{I_1(2 t)}{t} \,\,\,, 
\ee
with $I_1$ the modified Bessel function of first kind. Now the distribution governing
the probability that $\delta_{\cal O}$ is less than a value $M$ is given by 
$$ 
\Prob(\delta_{\cal O} < M) = \prod_{a = 1}^N \Prob( {\cal O}_{aa} < M) 
\simeq \left(1 - e^{- N I(M)} \right)^N \simeq 1 - N e^{- N I(M)}  \,\,\,. 
$$
We can find the scaling of the typical value of the maximum by requiring that the probability 
is large enough 
$$ 
\Prob(\delta_{\cal O} < M) \simeq \text{const.} \quad\Rightarrow\quad I(M) 
\simeq \frac{\ln N}{N} \,\,\,.
$$
Since $I(x) \xrightarrow{x\to0} 0$, we are interested in the behavior of $I(x)$ for small $x$ 
and from Eq. (\ref{Idef}) and Eq. (\ref{lambdadef}) we get
\be
\label{maxminscaling}
 I(x) \simeq \frac{x^2}{2} + O(x^3) \quad \Rightarrow \quad \delta_{\mathcal O} \,\,\,
 \simeq \sqrt\frac{\ln N}{N} \,\,\,. 
\ee
In Fig. \ref{minmaxdense}, one sees the numerical agreement with our heuristic
argument.

Finally let's consider the behaviour of the difference between the diagonal and the microcanonical ensembles
with increasing matrix size.  To this end, we analyzed the difference
$\sigma = \sqrt{({\mathcal O}_\text{micro} - {\mathcal O}_\text{diag})^2}$ of the two ensembles for
each initial state $|\psi_0\rangle$.
Due to the broad energy distribution of the overlap of the initial states, their
intensive energy $\overline{e}=\langle \psi_0|H(-h)|\psi_0\rangle$ lies in a 
region, $[-0.5,0.5]$, smaller than the range of the post-quench energies $-2<e_a<2$.  $\sigma$
show the same behavior independent of the particular initial state $|\psi_0\rangle$
being considered, as can be argued by the constant IPR combined with a structureless 
EEV distribution.
As all initial states are equivalent, we focus our attention 
on those initial states belonging to a small energy window around $e=0$: 
the result is shown in Fig. \ref{centerscalingdense}. As can be seen, 
the difference between micro-canonical and diagonal ensemble rapidly goes to zero as a function of N.

In conclusion, quenches in dense random matrices are characterized by initial states with
large IPRs, EEV distributions of the post-quench eigenbasis with no
energy dependence and whose variance goes to 
zero exponentially with increasing system size. In this sense, their thermalization is trivial, 
as the spread of the micro-diagonal ensemble is governed by a distribution whose
support is increasingly localized near zero as system size grows.

\section{Thermalization in Sparse Random Matrices}
\label{sparsematrices}

\begin{figure}[t]
\centering
$\begin{array}{cc}
\includegraphics[width=0.4\textwidth]{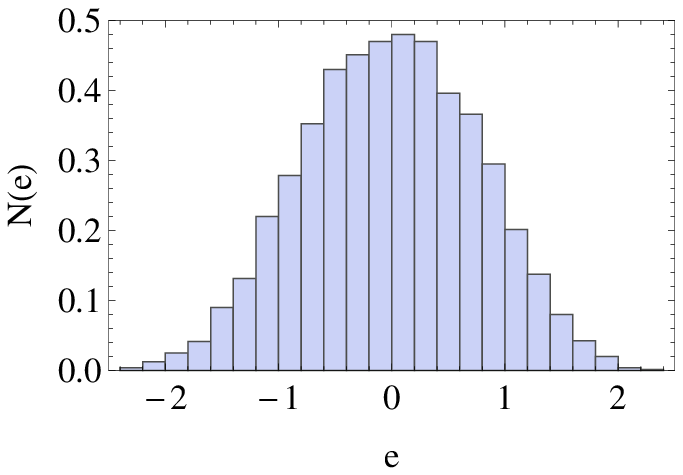} &
\includegraphics[width=0.4\textwidth]{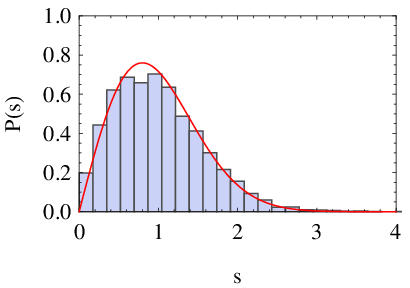}\\
\end{array}$
\caption{Sparse random matrices (with N=4000). Left: density of states, right:
level spacing statistics, the continuous line is the Wigner surmise for GOE
matrices.}
\label{dosspacingsparse}
\end{figure}

\begin{figure}[t]
\centering
$\begin{array}{cc}
\includegraphics[width=0.4\textwidth]{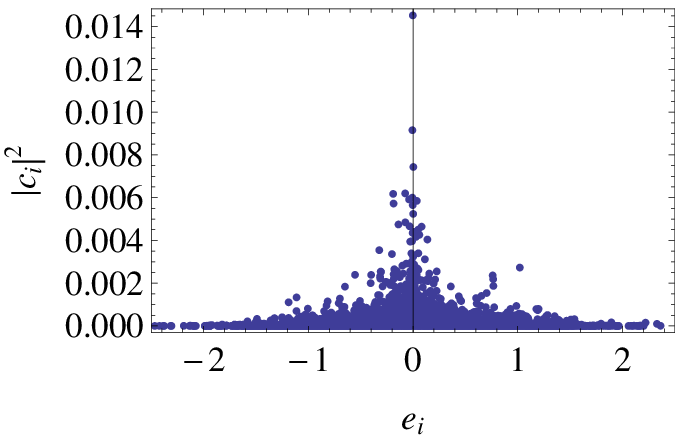}&
\includegraphics[width=0.4\textwidth]{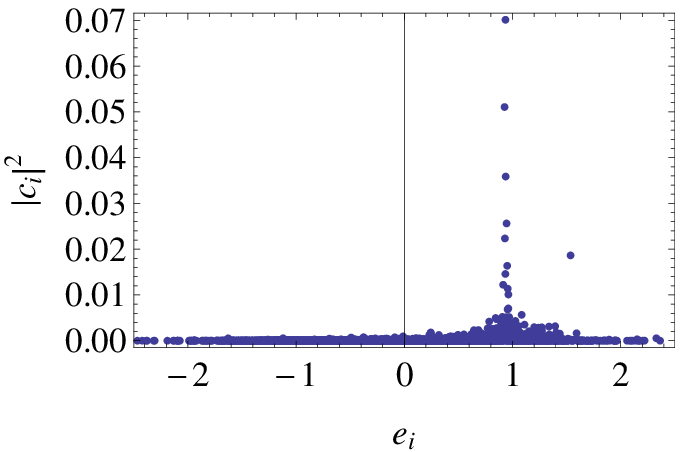}\\
\end{array}$
\caption{Sparse random matrices. Behavior of the overlaps $|c_a|^2$ for a state midspectrum (left) and for one 
in the upper portion of the spectrum (right).}
\label{sampleoversparse}
\end{figure}

\begin{figure}[b]
\centering
\includegraphics[width=0.4\textwidth]{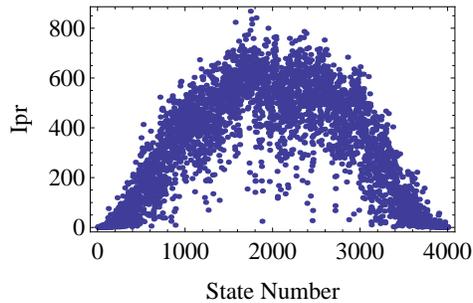} 
\caption{Sparse random matrices (with $N=4000$). The IPR for a specific realization.
The ordering of the initial states in this plot is according to their energy relative to the
post-quench Hamiltonian, $\langle \psi_0|H^>|\psi_0\rangle$.} 
\label{sampleiprsparse}
\end{figure}
\begin{figure}[t]
\centering
$\begin{array}{cc}
\includegraphics[width=0.4\textwidth]{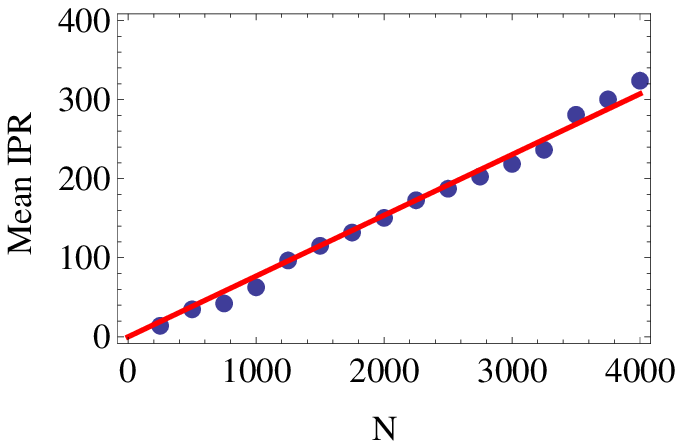} &
\includegraphics[width=0.4\textwidth]{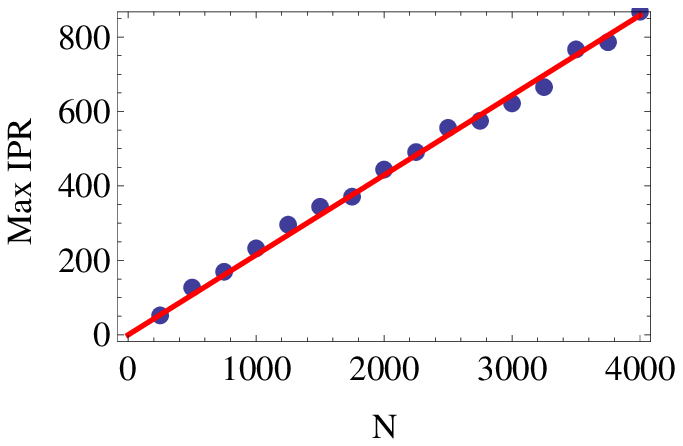}\\
\end{array}$
\caption{Sparse random matrices. IPR vs. matrix size. The continuous line is the $y=ax$ fit. Left, average IPR on the whole spectrum ($a=0.0738$). Right, maximum IPR ($a=0.2096$). }
\label{iprscalingsparse}
\end{figure}

We now turn to the more interesting case of thermalization in sparse random matrices.  As we have indicated
such matrices describe the Hamiltonians of systems with local interactions.
In order to define the ensemble of these matrices, we employed a symmetric mask matrix, $\mathcal{M}$.
This matrix has $1$'s on the diagonal and in each of its rows 
we allow it to have on average $\ln N$ off-diagonal entries equal to $1$.  All
remaining entries 
of the mask matrix are 
equal to zero.  
See Appendix B for details of how $\mathcal{M}$ is defined.
The upper triangular part of the (symmetric) Hamiltonian is then obtained as:
\be
\label{hamiltonianScaling}
H(h)_{i<j} = 
\left\{ \begin{array}{ll}
d_i ~{\rm with~}d_{i}{\rm~drawn~from~}\mathcal{N}(0,\ln N) & {\rm if}~i=j;\\
o_{ij}\times\mathcal{M}_{ij} ~{\rm with~}o_{ij}{\rm~drawn~from~}\mathcal{N}(0,1) & {\rm if}~i<j,
\end{array}\right.
\ee
where $\mathcal{N}(\mu,\sigma^2 )$ is a normal distribution with $\mu$ mean and $\sigma^2$ variance.
Then the coefficients in the off-diagonal blocks are multiplied times $h$ to reproduce the structure in Eq. (\ref{hamiltonianForm}).
The different choice for the variances of the diagonal $d_i$ and off-diagonal elements $o_{ij}$ 
is motivated by the requirement that the spectrum be extensive. 
\begin{figure}[b]
\centering
\includegraphics[width=0.4\textwidth]{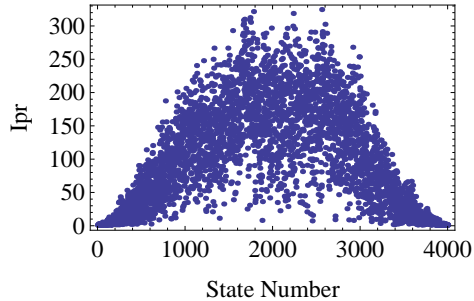} 
\caption{Sparse random matrices (with $N=4000$). The IPR  of the post-quench eigenstates w.r.t. the local basis for a specific realization.
The ordering of the initial states in this plot is according to their energy eigenvalue.} 
\label{sampleiprsparsewrtlocal}
\end{figure}
\begin{figure}[t]
\centering
$\begin{array}{cc}
\includegraphics[width=0.4\textwidth]{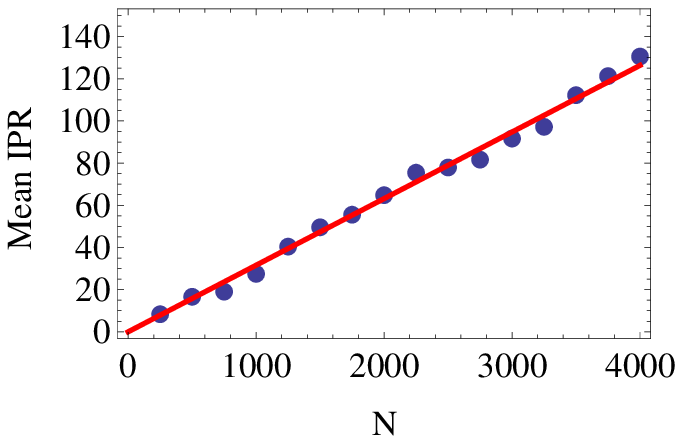} &
\includegraphics[width=0.4\textwidth]{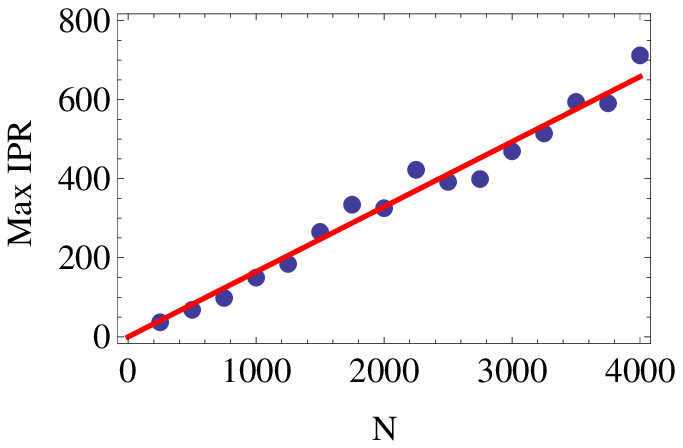}\\
\end{array}$
\caption{Sparse random matrices. IPR w.r.t. local basis vs. matrix size. The continuous line is the $y=ax$ fit. Left, average IPR on the whole spectrum ($a=0.0321$). Right, maximum IPR ($a=0.1632$). }
\label{iprscalingsparsewrtlocal}
\end{figure}
In fact we can
compute the variance of the spectrum:
\be
\label{sparsevariance}
\sigma_H^2 \equiv \frac{\Tr {H^2}}{N} \simeq 2 \ln N .
\ee
In the approximation in which the eigenvalues $E_a$ are considered independent
and normally distributed, one can relate the
minimum of the spectrum with the variance, obtaining for the groundstate energy
the estimate
$$
E_\text{gs} = \min_a \{E_a \} \simeq -\sigma_H \sqrt{2 \ln N} \simeq  - 2 \ln N .
$$
The eigenvalues midspectrum will be at most a few standard deviations $\sigma_H$ from the average $0$. 
Thus a typical eigenstate will be such that
\be
E_a^{(0)}-E_\text{gs}^{(0)} \propto \ln{N} \,\,\,.
\ee
We see then that our choices satisfy the requirement that energy is an extensive quantity.

We compare these estimates with numerics in 
Fig. \ref{dosspacingsparse} where we plot the density of states and the level
spacing distribution
for one realization of a sparse matrix. 
The density of states is no longer a semicircle, looking rather more like a bell-shaped distribution. 
Moreover, 
notice that, unlike the GOE case where the intensive quantities have a finite distribution 
in the large $N$ limit, i.e.Eq. (\ref{semicircleintensive}), 
in this sparse case the (intensive) standard deviation
behaves as $\frac{1}{\sqrt{\ln N }}$, while 
the support of the spectrum remains approximately $[-2,2]$.
We also see from the right side of Fig. \ref{dosspacingsparse} that the
level-spacing distribution obeys the Wigner form for 
a non-integrable model.

To generate the observables we follow a similar procedure, i.e. we employ the same mask $\mathcal{M}$. 
The idea behind this choice is that the matrix $\mathcal{M}$ is responsible for the local structure 
in the Hilbert space and so we keep it for all the physical observables (including the Hamiltonian itself). 
So we have for the upper triangular part of a symmetric observable 
$$ 
\mathcal{O}_{i<j} = \left\{ \begin{array}{ll}
                                    d_i {\rm ~with~} d_i {\rm~drawn~from~}
\mathcal{N}(0,1/\ln N) & i = j;\\
                                    o_{ij}\times M_{ij} {\rm ~with~} o_{ij}
{\rm~drawn~from~} \mathcal{N}(0,1/(\ln N)^2) & i\neq j.
                                   \end{array}\right.
$$
The even and odd parts are then obtained as before 
by splitting $\mathcal{O}$ into diagonal and off-diagonal blocks.

\begin{figure}[b]
\centering
$\begin{array}{cc}
\includegraphics[width=0.4\textwidth]{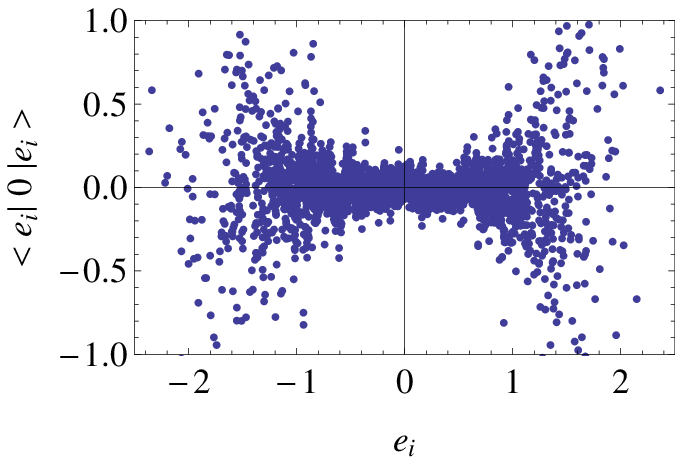} &
\includegraphics[width=0.4\textwidth]{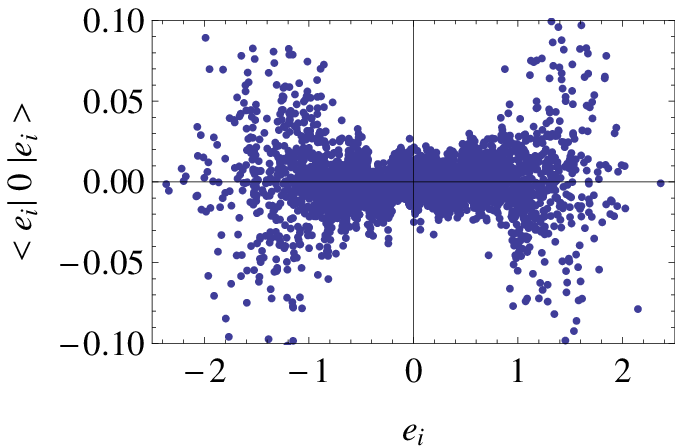}\\
\end{array}$
\caption{Sparse random matrices. EEV $\langle E_a|{\mathcal O}|E_a\rangle$ vs
$e_a$. Left: even observable. Right: odd observable}
\label{EEVsamplesparse}
\end{figure}
\subsection{Numerical results}
Unlike the dense case, in sparse random matrices the overlap distributions are peaked around the 
initial state post-quench intensive energy $\overline{e}$, as can be seen in two examples
shown in Fig. \ref{sampleoversparse}.

The IPR is no longer constant (as it was for the random dense matrices), but shows behaviour 
dependent on the energy of the initial state,  as demonstrated in
Fig. \ref{sampleiprsparse}.
However there is still scaling of the IPR with the matrix size, as can be seen by
studying the behaviour of the maximum IPR vs the matrix size, plotted in 
Fig. \ref{iprscalingsparse}.  In this case the mean and the maximum IPR are
rather different, 
due to the presence of states with very small IPR.

We see a similar phenomena when we study the IPR of the pre-quench and
post-quench 
(they are statistically equivalent) eigenbasis relative
to the local basis (that is, the basis of states in which the Hamiltonian matrices, $H(h^<)$ and $H(h^>)$
are expressed).  We see in Fig.\ref{iprscalingsparsewrtlocal} that the mean and
max IPR's here as a
function of matrix
size, $N$, are similar to those in Fig. \ref{iprscalingsparse}. 

\begin{figure}[t]
\centering
$\begin{array}{cc}
\includegraphics[width=0.4\textwidth]{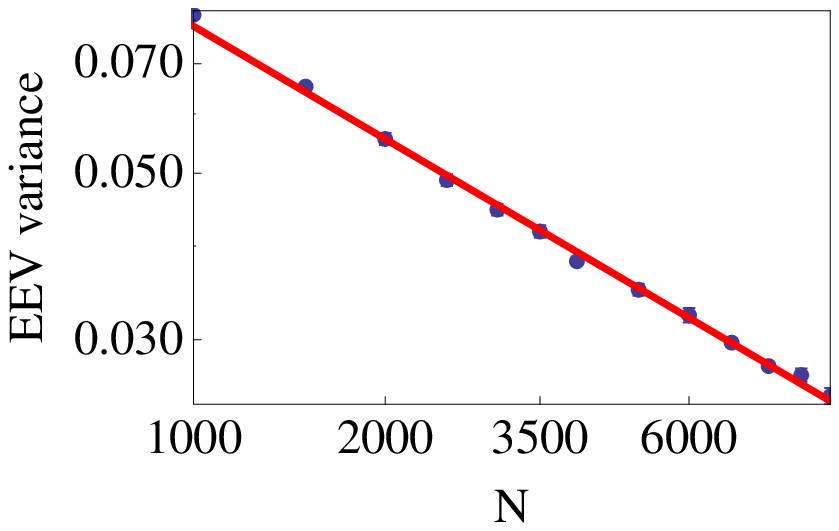} &
\includegraphics[width=0.4\textwidth]{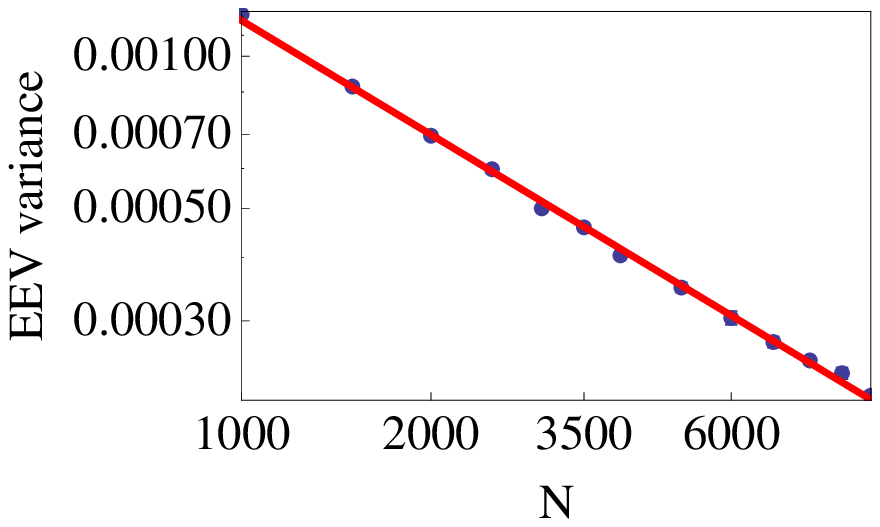}\\
\end{array}$
\caption{Sparse random matrices. EEV variance vs. matrix size for the full spectrum. The continuous lines are
the fit $a/x^b$. Left: even observable $b=0.50\pm 0.01$ Right: odd observable $b=0.75\pm 0.01$. }
\label{EEVvariancesparse}
\end{figure}
\begin{figure}[b]
\centering
$\begin{array}{cc}
\includegraphics[width=0.4\textwidth]{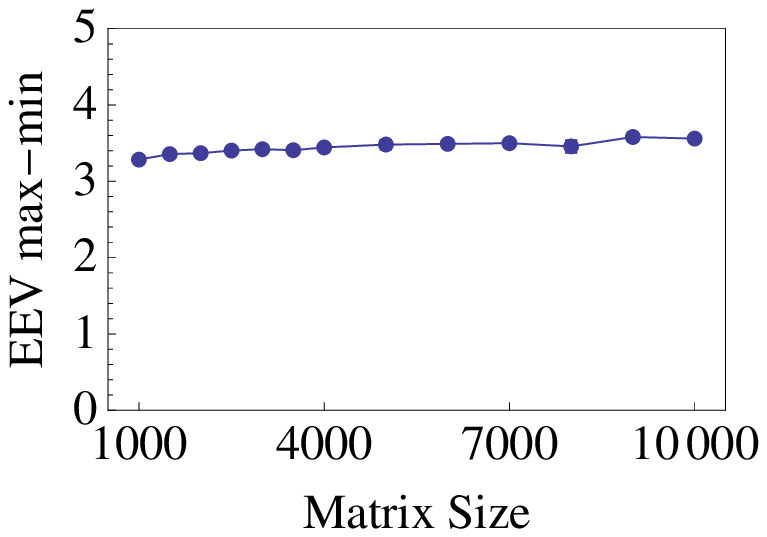} &
\includegraphics[width=0.4\textwidth]{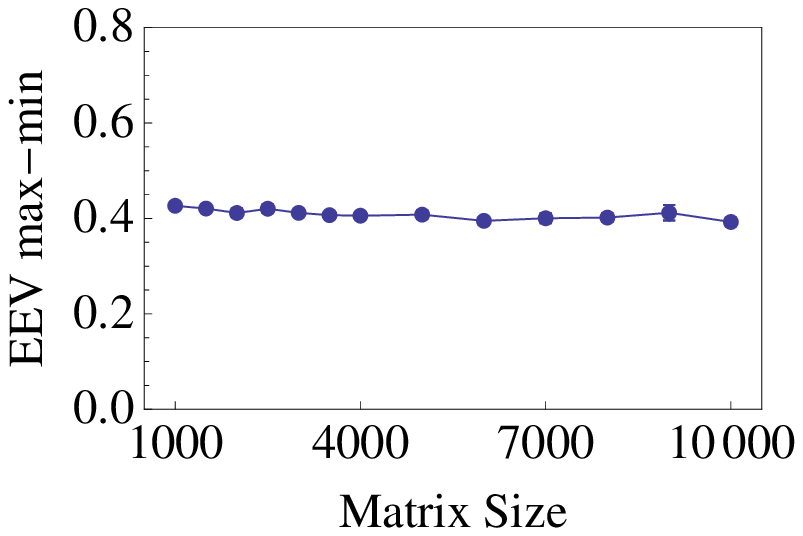}\\
\end{array}$
\caption{Sparse random matrices. Max-min EEV vs matrix size for the full spectrum. Left: even observable.
Right: odd observable.}
\label{minmaxsparse}
\end{figure} 

The equivalency of these two different IPRs is not surprising.
If we pick an eigenstate of the pre-quench Hamiltonian, $|\psi_0\rangle$, that has a small IPR relative to the
local basis, it will necessarily be only weakly coupled to the off-diagonal blocks, particularly as the Hamiltonian
matrices are sparse.  $|\psi_0\rangle$ will then be closely related to some post-quench eigenstate.
This implies in turn that $|\psi_0\rangle$ will have a small IPR relative to the post-quench eigenbasis.
Similarly a pre-quench eigenstate $|\psi_0\rangle$ with a large IPR in terms of the local basis will be strongly
affected by the quench in the sense that it is unrelated to any eigenstate of the post-quench eigenbasis,
and so will have a large IPR in terms of this basis.  It is this that lies behind the similar shapes
of Fig. \ref{sampleiprsparsewrtlocal} and Fig. \ref{sampleiprsparse}.

The behavior of the IPR relative to the local basis (and, by this equivalency, the IPR relative to the post-quench eigenbasis)
can be understood directly in the framework of Anderson localization \cite{Anderson}.  Even though our
problem is a many body one, our 
sparse Hamiltonian can be seen as akin to the dynamics of a non-interacting particle hopping
on a Bethe-lattice of fixed 
connectivity, $\ln N$, where each site has a random potential (the diagonal part of the random Hamiltonian).
It is well known \cite{Abouchacra} that for this model the Anderson localization transition occurs 
with the presence of a \textit{mobility edge} which separates the delocalized states (in the middle 
of the band) from the localized states (in the tails of the energy spectrum).
In our case localization occurs 
when the $IPR$ is $O(1)$, while delocalization is seen for $IPR = O(N)$.

\begin{figure}[t]
\centering
$\begin{array}{cc}
\includegraphics[width=0.4\textwidth]{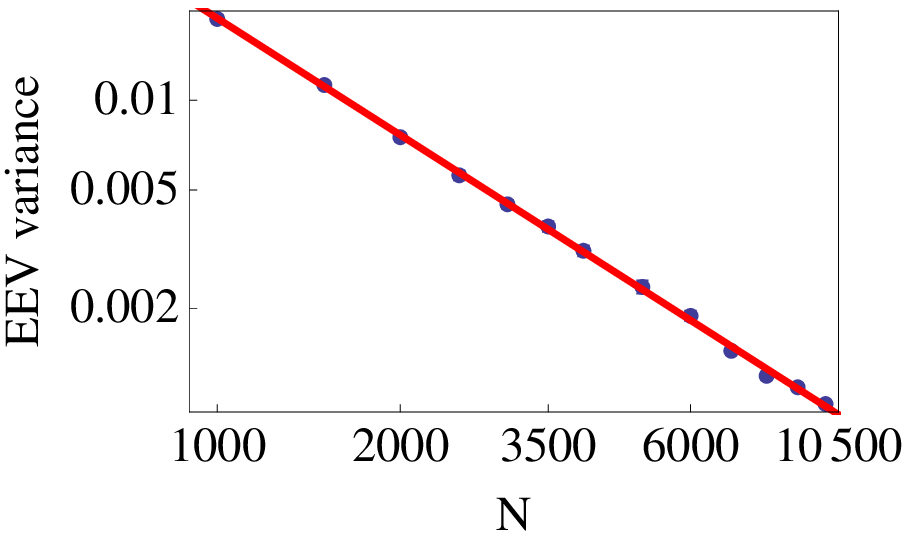} &
\includegraphics[width=0.4\textwidth]{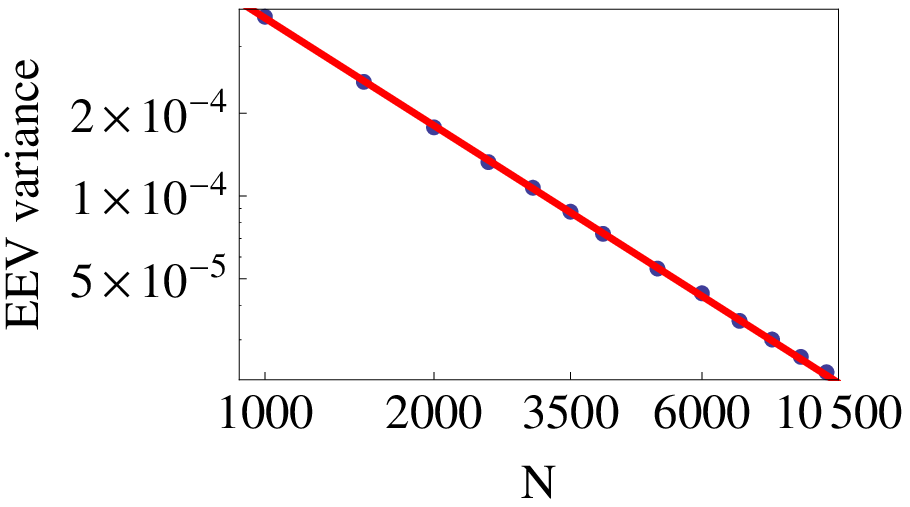}\\
\end{array}$
\caption{Sparse random matrices. EEV variance vs. matrix size for a small energy window around $e=0$. 
The continuous line is the fit $a/x^b$. Left: even observable $b=1.29\pm 0.01$.  Right: odd observable $b=1.2\pm 0.1$.}
\label{EEVvariancesparsecenter}
\end{figure}
\begin{figure}[b]
\centering
$\begin{array}{cc}
\includegraphics[width=0.4\textwidth]{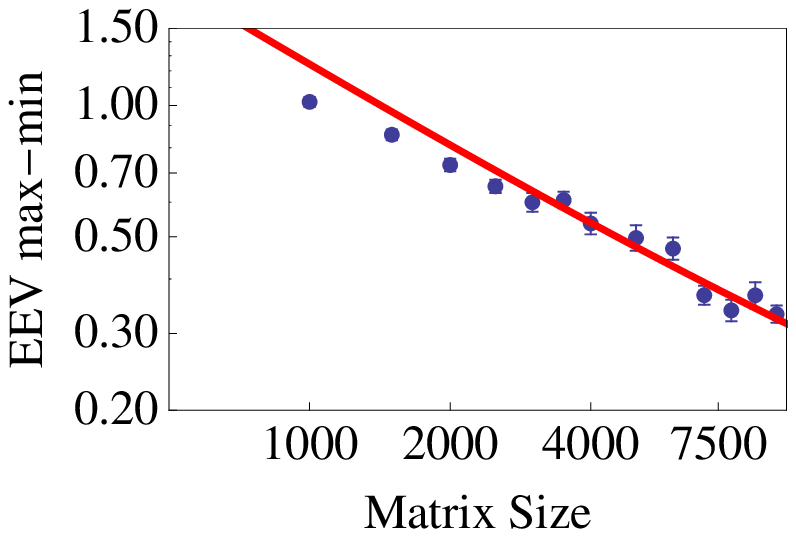} &
\includegraphics[width=0.4\textwidth]{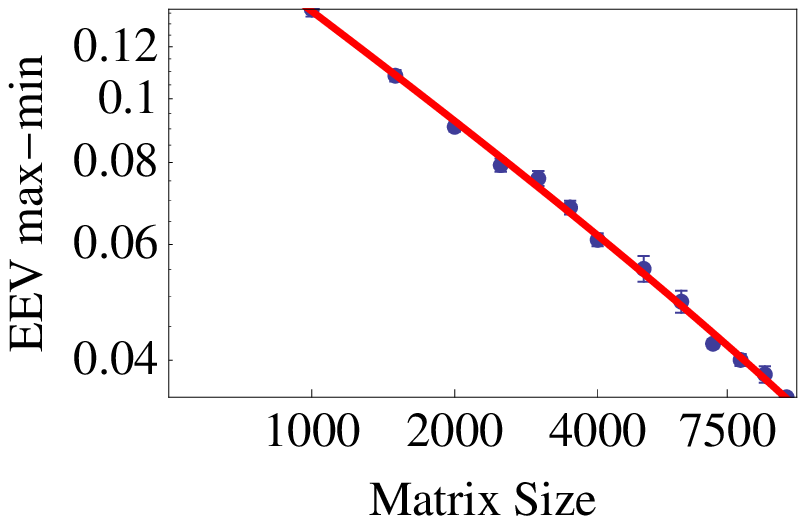}\\
\end{array}$
\caption{Sparse random matrices. Max-min EEV vs. matrix size  for a small energy window around $e=0$ in logscale. The continuous line is the fit $a/x^b+c$. Left: even observable $e=0$, $b=0.65\pm 0.01$, $c=0.067\pm 0.001$. Right: odd observable $e=0$, $b=0.50\pm 0.01$, $c=0.01\pm 0.01$.}
\label{minmaxsparsecenter}
\end{figure}

\begin{figure}[b]
\centering
$\begin{array}{cc}
\includegraphics[width=0.4\textwidth]{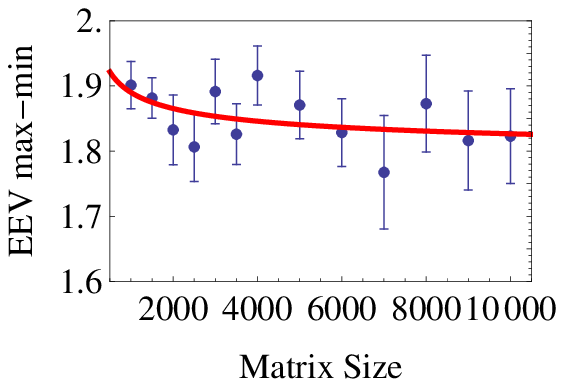} &
\includegraphics[width=0.4\textwidth]{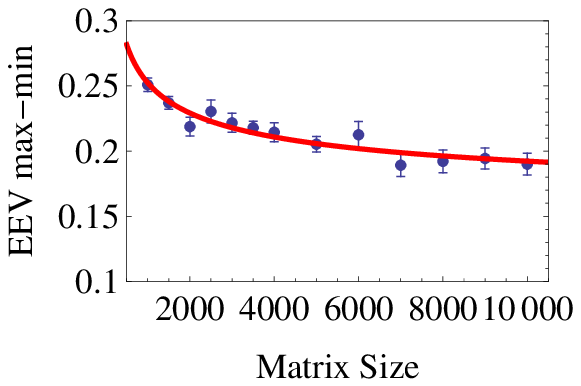}\\
\end{array}$
\caption{Sparse random matrices. Max-min EEV vs. matrix size  for a small energy window around $e=1$. The continuous line is the fit $a/x^b+c$. Left: even observable $e=1$, $b=0.35\pm 0.01$, $c=1.7\pm 0.1$. Right: odd observable $e=1$, $b=0.33\pm 0.01$, $c=0.14\pm 0.01$.}
\label{minmaxsparseone}
\end{figure}

The position, $E_m$, of the mobility edge can be 
determined by the following equation, derived along the same lines as \cite{Abouchacra}
%(where 
%we have taken the connectivity $K$ to be $\ln N$, the hopping term $V$ to be $1$, and the 
%function $Q(x)$ to be $p(R-x)$, given below).  {\color{red} What is R?}
%With this substitutions, the equation
\be
\label{aboumobility}
2\ln N \int_0^\infty \{ p'(x-E_m) - p'(E_m-x) \} \ln x \,dx= 1 \,\,\,, 
\ee
where $p(x)$ is the probability density of the diagonal terms in our Hamiltonian
\be
\label{gaussianDistr}
 p(x) = \frac{1}{\sqrt{2\pi \ln N}} \exp \left(- \frac{x^2}{2 \ln N} \right) \,\,\,. 
\ee
States with an $|E| > |E_m|$ such that the right hand side of Eq. (\ref{aboumobility}) is less than $1$ 
are localized.  Otherwise they are delocalized.
The integral in Eq. (\ref{aboumobility}) can be estimated at the leading order in the large $N$ limit, and
one obtains
\be
\label{mobility}
1 =  \exp\left(-\frac{E_m^2}{2\ln N}\right) \ln\ln N \sqrt{\frac{2\ln N}{\pi}}
\quad\Rightarrow \quad E_m \simeq \pm \sqrt{\ln N \ln \ln N} \,\,\,. 
\ee
Thus in the large $N$ limit all the states with a non-zero intensive energy $E/\ln N$ behave 
as localized. Nonetheless the majority of the states, as they are concentrated in a window
of width $\sigma_H$ (given in Eq. \ref{sparsevariance}) and as  $E_m/\sigma_H \gg 1$, will be delocalized with an $IPR = O(N)$.

%It is worth however stressing the slowness of 
%the scaling as a function of $N$ in the quantity above:  taking $L \simeq 10^{23}$, one has 
%$$ 
%\ln\ln N \simeq \ln L \simeq 53,
%$$ 
%and so therefore even in this case
%a finite fraction of the spectrum remains localized.

\begin{figure}[t]
\centering
$\begin{array}{cc}
\includegraphics[width=0.4\textwidth]{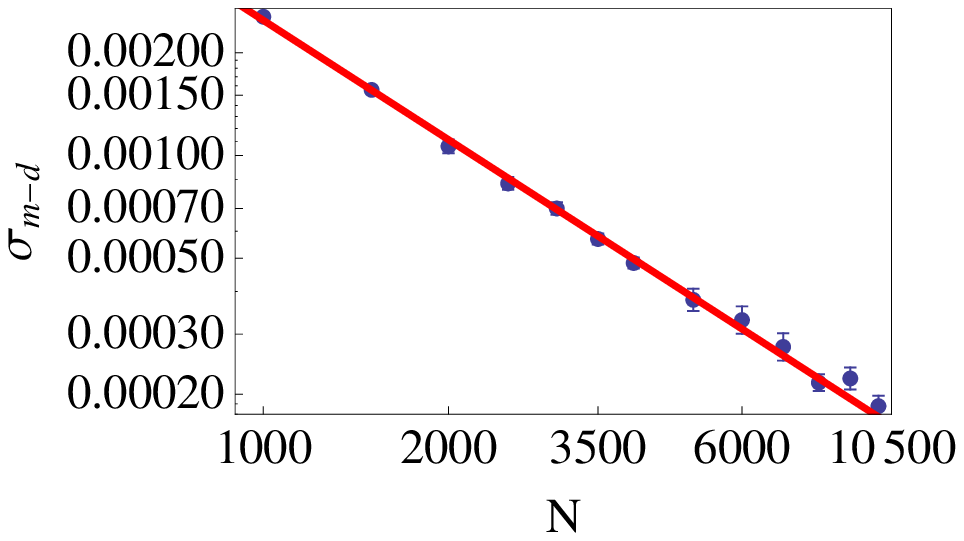} &
\includegraphics[width=0.4\textwidth]{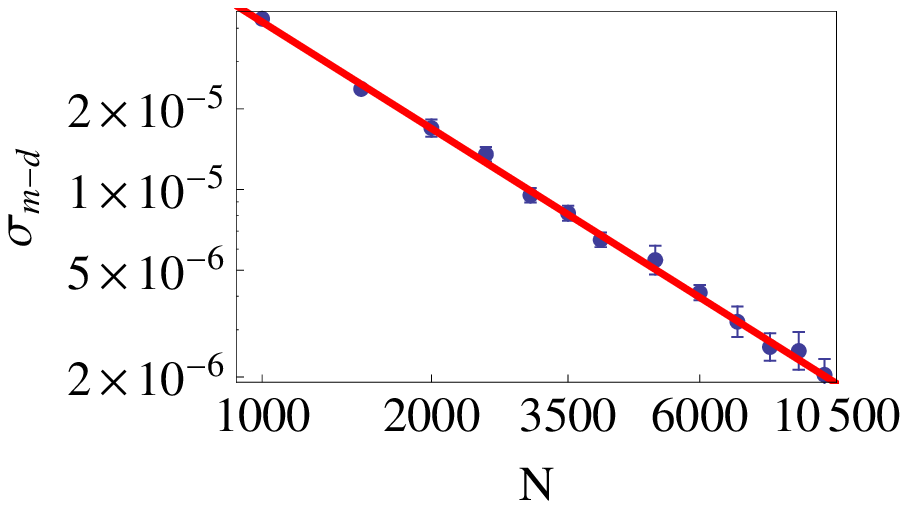}\\
\end{array}$
\caption{Sparse random matrices. $\sigma=\sqrt{({\mathcal O}_\text{micro} - 
{\mathcal O}_\text{diag})^2}$ vs. matrix size for initial states laying in the central part of the 
spectrum $\overline{e}\approx 0$. The continuous lines are the fits $a/x^b$. Left: even observable $b=1.15\pm 0.01$.
Right: odd observable $b=1.30\pm 0.01$.}
\label{centerscalingsparse}
\end{figure}

\begin{figure}[t]
\centering
$\begin{array}{cc}
\includegraphics[width=0.4\textwidth]{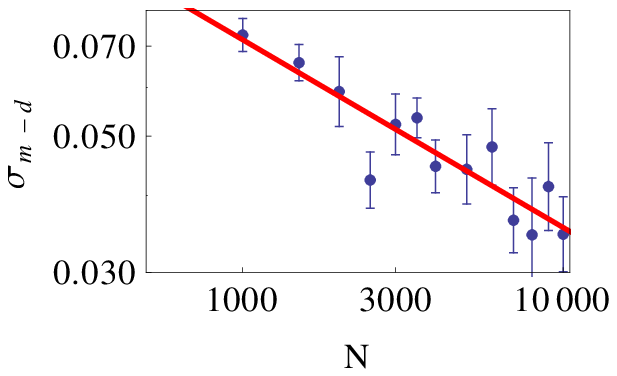} &
\includegraphics[width=0.4\textwidth]{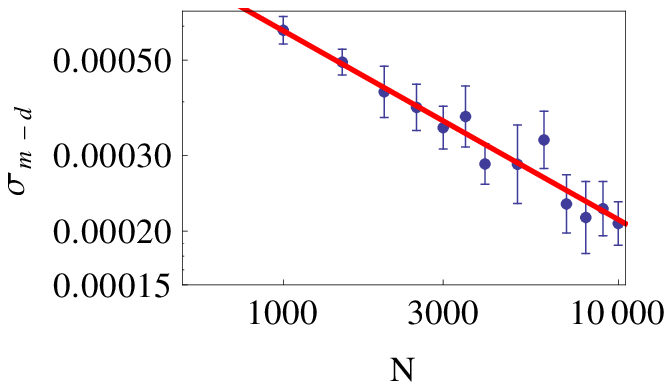}\\
\end{array}$
\caption{Sparse random matrices. $\sigma=\sqrt{({\mathcal O}_\text{micro} - 
{\mathcal O}_\text{diag})^2}$ vs. matrix size for initial states laying in a small window around $e=1$. The continuous lines are the fits $a/x^b$. Left: even observable $b=0.31\pm 0.01$.
Right: odd observable $b=0.44 \pm 0.01$.}
\label{centerscalingsparsee1}
\end{figure}

\begin{figure}[b]
\centering
\includegraphics[width=0.4\textwidth]{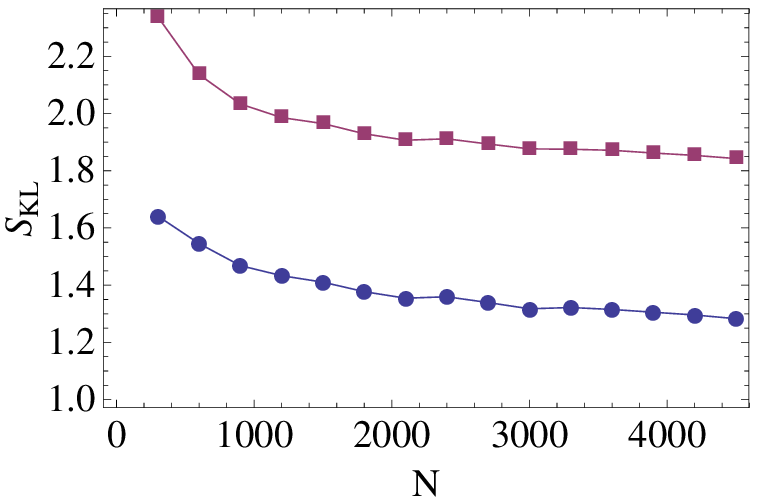}
\caption{Sparse random matrices. Kullback-Leibler entropy vs matrix size for the center of the 
spectrum for the uniform (squares) and the Gaussian distribution (circles). }
\label{KLscaling}
\end{figure}
The localization of states with finite intensive energies has implications for the EEV distribution
vs. $e_a$.  As the observables have a matrix structure closely related to that of the Hamiltonians,
we expect that localized states to be close to eigenstates of the observables itself, in contrast 
with Eq. (\ref{overlapIPR}) in the dense case.
On such states, the observables will have expectation values far from their zero
average.  On the contrary, the delocalized states midspectrum will have EEV values closer to the mean of zero.
In Fig. \ref{EEVsamplesparse} we see numerical verification of this.

This result marks a strong difference with respect to the case of dense matrices.  We also see 
marked differences between the sparse and dense cases with both the variance and the support of 
the EEVs distribution of the observable, 
as shown in Fig. \ref{EEVvariancesparse} and \ref{minmaxsparse}: while the variance 
approaches zero as $N$ grows, the support does not, instead tending to a non-zero constant. 
Therefore the scaling Eq. (\ref{maxminscaling}) is no longer applicable, most likely
as the overlaps, $A_{i\beta}$,
and the eigenvalues of the observable, $O_\beta$, in Eq. (\ref{observableexpansion}) can no longer
be considered as independent.

Since the distribution of the overlap coefficients, $c_a$, are peaked around the energy 
$\overline{e}=\langle \psi_0 |H(h)|\psi_0\rangle$, it is worthwhile
to analyze the scaling
behavior of the distribution of the EEVs in the vicinity of a specific $\bar e$. 
Here we choose two different energy windows, one centered around $\bar e = 0$, lying exactly mid-spectrum and one around $\bar e = 1$. For  $\bar e = 0$ 
the variance and the max-min spread
as functions of the size $N$ are plotted in Fig. (\ref{EEVvariancesparsecenter}) 
and Fig. (\ref{minmaxsparsecenter}).  It is this distribution
that is going to determine whether with respect to a particular observable we see thermalization.
We again see that the variance is going to zero, 
while in contrast to the full spectrum, the max-min spread seems to tend, asymptotically, 
to either a very small constant value or to zero.  The errors in our numerics are then not small enough to tell us at $\bar e = 0$
whether there is a complete absence of rare states.
However we can be more definitive for the energy window centered at $\bar e = 1$.
In this window we see (as evidenced in Fig. \ref{minmaxsparseone})
that the max-min spread of the EEV's in the large N limit tends to a finite constant for both the even and 
odd observables.

Our numerical data then suggests that there are rare states where the observable remains far from its average value
in each microcanonical
energy window corresponding to non-zero intensive energy, 
differently for what has been observed in the $t-J$ model \cite{sr1,sr2,sr3}. 
In these studies of the t-J model, rare states were argued to be absent 
for a range of strengths of the next nearest neighbour interaction, $V'$, and for an energy window centered mid-spectrum.  
In particular with $V'$ small the t-J model is effectively
integrable and rare states were found to be present and with $V'$ large, the model develops energy bands, also compatible with the
existence of rare states.  It is for a middle range of $V'$ that rare states were then found to be 
absent.  Our own study sees some agreement with these results.  And it is natural to think
there would be some agreement at least.  Our study of random matrix model should correspond to the $t-J$ model with intermediate
values of $V'$:  our matrix models are neither integrable nor do they have any notion of energy bands.
In our case, rare states (at least for what we call the odd observable) seem to be vanishing
as system size increases exactly in the center of the spectrum.  However away from this midpoint of
the spectrum, we find that rare states do exist, even in the thermodynamic limit.  It would thus be interesting
to extend the work of \cite{sr1,sr2,sr3} to additional energy windows.

Following \cite{Biroli}, we then conclude in the case of sparse matrices
that thermalization may depend on the particular nature of the
initial state and will not occur when such rare states are given a proportionally
large weight in the decomposition of the initial state.
We, however, do not find for the particular initial conditions specified by our quench protocol that 
the rare states are given disproportional weight such that thermalization does not occur.
For both energy windows $\bar e =0$ (see Fig. (\ref{centerscalingsparse})) and $\bar e=1$ (Fig. (\ref{centerscalingsparsee1})),
we see that with increasing matrix size the difference 
between the diagonal and microcanonical ensembles averaged over all initial conditions tends to zero.
This implies that the weighting of rare states in our initial states is not preponderant.  We do note however that 
the vanishing of the difference between ensembles decreases considerably more slowly with system size for the energy
window, $\bar e =1$, than for $\bar e =0$.  We might ascribe this to the presence of rare states at this energy -- even though these states
do not lead to non-thermalization in the thermodynamic limit, they may slow the approach to a thermalized state
as the system size is increased.

We verify this by computing the Kullback-Leibler entropy.  This entropy
is an information theoretic tool used to estimate how close two distributions are. It is
defined as
\be
S_{KL}=\sum_a P(a)\ln{\frac{P(a)}{Q(a)}} \,\,\,, 
\ee
where $P(a)$ is the expected distribution and $Q(a)$ is the distribution to be
compared. $S_{KL}$ is zero if the two distributions coincide except for sets of
zero measures. In our case we choose $P(a)=|c_a|^2$ and $Q(a)$ to belong to either 
a uniform or a Gaussian distribution centered about the energy $\overline{e}$. The 
range over which $Q(a)$ is defined has been taken such that the variances of $Q$ and $P$ coincides. 
Fig. \ref{KLscaling} shows the average KL-entropy vs. matrix size for the central
part of the spectrum which indicates in both cases a slow decay
as the system size is increased.  
The slow approach of the distribution $P(a)$ to a distribution $Q(a)$ 
that is both smooth and symmetric about $\bar e$
suggests that rare states are not weighted in a peculiar way, thus permitting thermalization.

In conclusion, for sparse random matrices our numerical data is compatible with the existence of rare states. 
However, the initial states selected by the quench protocol do not seem to have large overlaps with these
rare states and so we typically find thermalization as the end result of our quench process. 

\section{Time Scales of Thermalization}
\label{timescales}
\begin{figure}[t]
\centering
$\begin{array}{cc}
\includegraphics[width=0.4\textwidth]{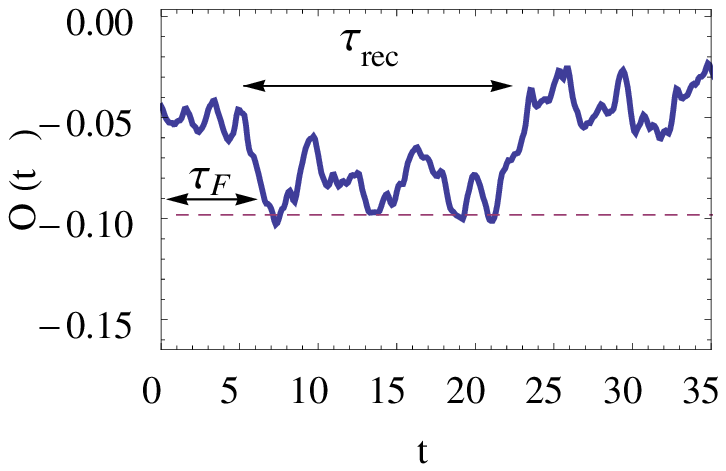} &
\includegraphics[width=0.4\textwidth]{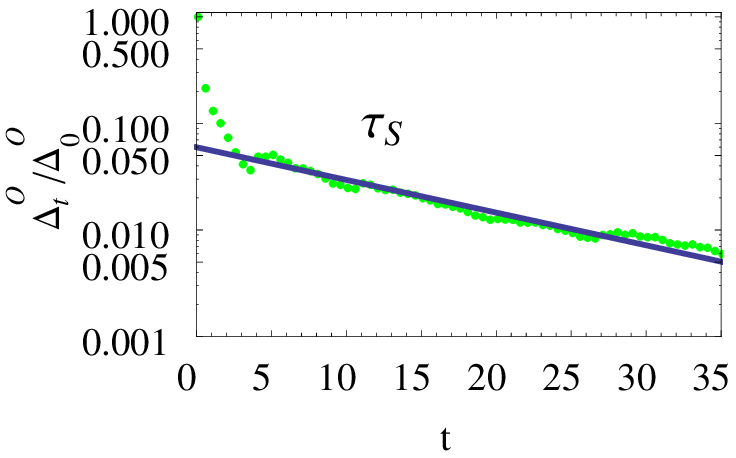}\\
\end{array}$
\caption{Left: time evolution of the observable $\mathcal{O}_t$ versus time $t$. 
Right: time evolution of the fluctuations $\Delta_t^{\mathcal{O}}/\Delta_0^{\mathcal{O}}$ versus time $t$, logscale on $y$ axis. The straight line is the exponential fit whose slope defines $
\tau_S$.} 
\label{timeEvol}
\end{figure}
The sparse random ensemble, inasmuch as it mimics some characteristic properties of thermalization in systems with
local Hamiltonians, is the right framework to address the study of the thermalization time. 
Interest in this quantity can be traced back to the seminal paper by Von Neumann \cite{vonNeumann} regarding 
the quantum ergodic theorem (QET).  The statement made in von Neumann's paper is that, under suitable 
assumptions (the Hamiltonian has no resonances - meaning that the energy level differences are non degenerate),
any state $\ket{\psi_0}$ in the energy shell $[e-\Delta, e+\Delta]$, will thermalize for most choices of the 
observable and most times $t$, i.e.
$$ 
\Ot \,=\, \bra{\psi_0(t)} \mathcal{O} \ket{\psi_0 (t)} = \Tr{\mathcal{O} \rho_{\text{mc}}} 
\qquad \text{for almost all } t, \mathcal{O} \,\,\,. 
$$
To make the notion of (macroscopic) observable and ``most'' used here precise would require the
development of an involved technical apparatus and so instead
we refer the reader to the existing literature \cite{vonNeumann, Goldstein}.
Nonetheless we can say there are important differences between this quantum thermalization and 
the classical notion of ergodicity where a time-average is involved. 
It is therefore of interest to give a precise estimate of the time needed for thermalization.
There have been different approaches which have tried to clarify this question
\cite{Vinayak,Masanes,Brandao,Short}. 
In Fig. \ref{timeEvol}, we examine the 
typical behavior of a realization of a random observable, $\Ot$.  
It decays towards the average value given by the diagonal ensemble:
$$ 
\mathcal{O}_\infty = \Tr{\mathcal{O} \rho_{\text{diag}}} \,\,\,,
$$
and we define the time $\tau_F$ as the first time at which $\mathcal{O}_t$ meets $\mathcal{O}_\infty$.
\begin{figure}[b]
\centering
$\begin{array}{cc}
\includegraphics[width=0.4\textwidth]{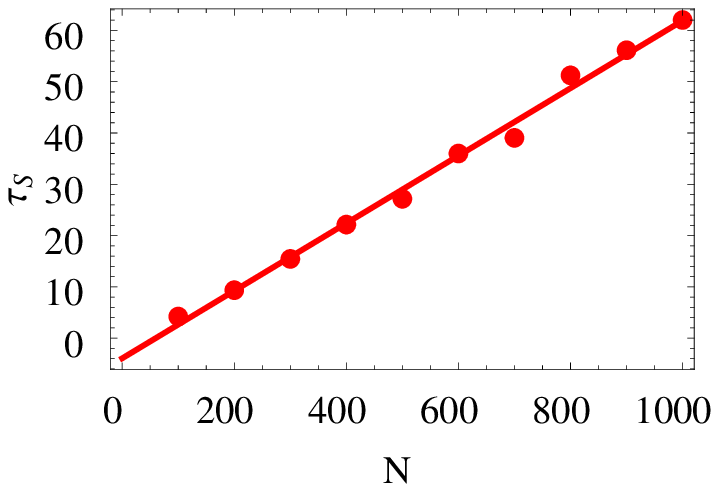} &
\includegraphics[width=0.4\textwidth]{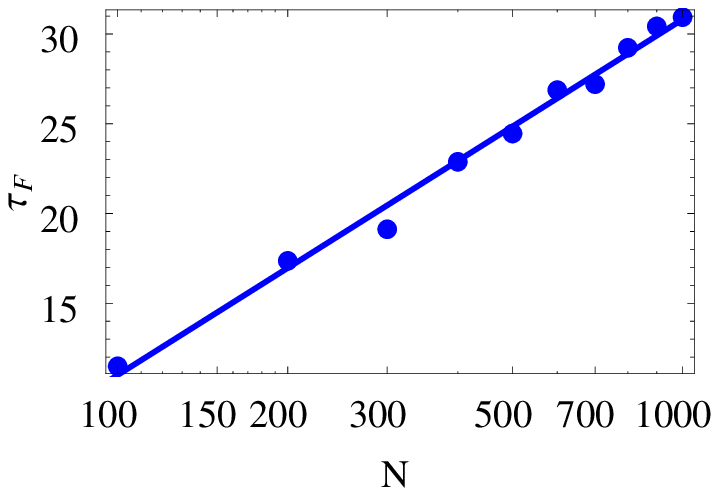}\\
\end{array}$
\caption{Scaling with the matrix size of the two time scale $\tau_S$ (left) with a linear fit ($-3.94 + 0.066 N$) and 
$\tau_F$ (right) with a logarithmic fit ($-28.61 + 8.60 \ln N$).}  
\label{timeScaling}
\end{figure} 
We stress here, that even if this quantity has not a direct physical interpretation, it can be considered as a lower bound for the thermalization time.
The time evolution of the observable however keeps fluctuating around the average, due to finite
system size, with it coming close to its initial value after a time, $\tau_{rec}$, the recurrence time.
To quantify these fluctuations we define
$$ 
\Delta^\mathcal{O}_t \equiv \frac{1}{t}\int_0^t ( \Ot - \mathcal{O}_\infty)^2 d\tau  
\xrightarrow{t\to\infty} \sum_{a,b} c_a^2 c_b^2 O_{ab}^2 \equiv \Delta^\mathcal{O}_\infty \,\,\,, 
$$
where we have assumed the absence of energy degeneracies and resonances. 
The quantity $\Delta^\mathcal{O}_t$ can be considered as the variance in the time interval $[0,t]$ of the observable 
and its behavior is plotted in Fig. \ref{timeEvol}: relaxation to the infinite time value is found as expected.
We can fit this curve supposing an exponential relaxation, $e^{- t/\tau_{S}}$, 
defining in this way another time scale, $\tau_{S}$, the time interval needed for the 
relaxation of the fluctuations. Notice that this quantity is the one closer to the Von Neumann formulation: 
indeed, from the Chebyshev inequality, one has a bound on the fraction, $\mu/t$,
of times where the observable has an 
expectation value far from its average
$$ 
\frac{\mu (\tau \in [0,t] \; {\rm ~and~} \; |{\cal O}_\tau - \mathcal{O}_\infty| > a)}{t} < 
\frac{\Delta^\mathcal{O}_t}{a^2} \,\,\, .
$$  
As in the thermodynamic limit $ \Delta^\mathcal{O}_\infty \to 0 $, this fraction must also go to zero
in the long time limit. In Fig. \ref{timeScaling}, we see the comparison between the 
two timescales $\tau_{S}$ and $\tau_{F}$ versus the system size. From this plot one can see clearly 
that $\tau_S$ is a long time-scale, with a behavior proportional to $N$, i.e.
the size of the Hilbert space.
Notice that this can also be interpreted as the
minimum spectral gap and at the leading order in $N$:
$$ 
\min_{a\neq b} |E_a - E_b| \propto \frac{1}{N}. 
$$
In contrast, $\tau_F$ is characterized by a much slower scaling with the size of the system and is 
therefore a fast time-scale. Although it is not easy to 
extract the precise scaling law from the available data, we have fit this data with the form
$$ 
\tau_F(N) = a \ln N \,\,\, ,
$$
and so taking the scaling of this time scale to go as the volume.
In contrast, the time scale for dense matrices was recently argued to go as the {\it inverse} of the volume\cite{Brandao}.

\section{Conclusions}
\label{Conclusions}

In this paper we have addressed the issues of thermalization and the Eigenstate Thermalization Hypothesis in the framework of 
random matrices, aiming to identify  
certain statistical properties of quantum extended systems subjected to a quench 
process. For this purpose we focused our attention on  $Z_2$ breaking quantum 
Hamiltonians, among the simplest theoretical quench protocols. In an attempt to 
encode in our analysis the property of locality, we have considered the ensemble of 
sparse random matrices and we have compared the data coming from this ensemble with 
similar data extracted from the ensemble of dense random matrices. We have found reliable 
evidence of different behavior in the two ensembles.  These differences show up both in the 
IPR of the quench states and in the distribution of the expectation values of the observables on 
post-quench energy eigenstates. In particular, while in the dense random matrix ensemble both the 
variance and the support of the observables vanish with increasing system size $L$, 
the sparse random matrix ensemble sees instead strong indications that the variance of EEVs goes to zero while 
the support remains finite as $L \rightarrow \infty$. The different 
behavior of the two ensembles can be traced back to the different density of states exhibited by the 
two sets of matrices: while in the dense matrices all states are delocalized in the Hilbert space, 
with almost equal overlap on all energy eigenstates, in the sparse matrices there are instead both 
delocalized and localized states. Localized states give rise to rare values of the expectation values 
of the observables, i.e. values which differ from the typical ones sampled by the micro-canonical ensemble. 
If properly weighted, such localized states may give rise to a breaking of thermalization. In the absence 
of such weighting, as seems to be the case in the initial conditions chosen by our quench protocol,
one instead observes relaxation to the thermal value of the local observables. 

In the framework of sparse random matrices, we have also provided numerical
estimates of the different time scales of thermalization. We have found that it is
possible to identify two time scales: a fast one $\tau_F$ and a slow one $\tau_S$, and that they depend 
differently upon the size of the system. 

\vspace{5mm}
\noindent
{\bf \large Acknowledge}
\vspace{1mm}

\noindent
We are grateful to M. Rigol, L. Santos, P. Calabrese, V. Kravtsov and  M. M\"{u}eller for useful discussions. 
RMK acknowledges support by the US DOE under contract number DE-AC02-98 CH 10886. 

\newpage
\appendix 
\section{Total momentum and quasiparticles}
\label{appendixMomenta}
We want to consider in this appendix the effect of a change of basis on the sparseness 
of the Hamiltonian matrix. In particular we will focus on a change to a basis where momentum
is a good quantum number. We take once
again the Ising case as a guiding example. Changing the basis, the new matrix
representation of the Hamiltonian Eq. (\ref{Ising}) can be obtained in terms of an
unitary matrix $U$:
\begin{equation}
 \label{changeofbasis}
\mathcal{H}' \,=\, U^\dag \,\mathcal H \,U
\end{equation}
and, for a general change of basis, the resulting matrix will not be sparse
anymore; this is quite intuitive: for an arbitrary unitary matrix $U$, i.e. an arbitrary basis, any
notion of locality is lost. 

However we can ask what happens in a much more common case: is the Hamiltonian
still sparse in the basis of
momenta? We will show that there is a
subtle issue related to the definition of this change of basis. To be more
specific, let's put the Ising Hamiltonian on a circle by introducing periodic
boundary conditions such that the $(L+1)$-th site is identified with the first one. In
this case the system becomes translationally invariant; this can be formally
written as:
$$ [H, T] = 0, $$
where $T$ is the shift operator, defined by:
$$ T\ket{m_1,\ldots, m_L} = \ket{m_L, m_1, \ldots, m_{L-1}}. $$
Clearly we have:
\begin{equation}
 \label{nilpotent}
T^L = \mathbf{1},
\end{equation}
and therefore we can define the total momentum $P$ as:
$$ T = e^{ i P}  \Rightarrow [H,P] = 0. $$
As $T$ is a unitary operator, using Eq. (\ref{nilpotent}) we easily deduce the
usual structure of the spectrum of $P$ as $\frac{2\pi n}{L} $. The basis of
momenta is then the basis of the eigenstates of the total
momentum. However, the eigenvalues of $P$ are highly degenerate and so many definitions are
possible. Two examples will demonstrate that sparseness depends on which definition
is employed.

Let us first 
construct a complete set of eigenstates of $P$ in the following way. We first pick
a state $\ket s = \ket{m_1,\ldots m_L}$ in the real-space basis. We obtain
an invariant subspace for $T$ by considering the set of states:
$$ I_s = \operatorname{Span}\{\ket s, T \ket s, T^2 \ket s, \ldots , T^{L_s-1}
\ket s\} $$
where $L_s$ is been defined as:\footnote{The minimum exists because the set
contains at least the element, $L$.  We also note that $L_s$ divides $L$.}
$$L_s \equiv \min\{n {\rm~such~that~} T^n \ket s = \ket s \} $$
The eigenstates of $P$ composed of linear combination of states in the set $I_s$ can be written as:
\begin{equation}
 \label{fourier1}
\ket{\tilde s_n} = \sum_{k = 0}^{L_s-1} e^{\frac{2 \pi i n k}{L_s}} T^k\ket s
\quad \Rightarrow \quad P \ket{\tilde s_n} = \frac{2 \pi n}{L_s} \ket{\tilde s_n
}.
\end{equation}
A full basis of eigenstates for $P$ can be obtained repeating this procedure
for different states $\ket s$: we call this basis the \textit{rigid-translation
Fourier basis} (RTFB). It is easy to understand that the resulting
matrix in this new basis is still sparse. In fact each state is a superposition
of at most $L$ states and it follows the corresponding transformation $U$
contains at most $L$ non-zero entries in each row and column. Using
Eq. (\ref{changeofbasis}), we conclude that the Hamiltonian matrix in this basis is
still sparse having at most $L^3$ non-zero entries in each row ($L^3 \ll 2^L$).

Is this the end of the story? Let us consider what happens if we use
a different momentum basis.  To this end we work in a
second-quantization framework and focus upon a set of operators
satisfying (where the $\pm$ stands for the fermionic and bosonic case):
$$ 
[a_i, a^\dag_j ]_{\pm} = \delta_{ij} .
$$ 
These operators create and destroy a single particle excitation at position $i$. A real-space
basis can be written in this formalism as:
\begin{equation}
 \label{secondquantizationrealspace}
 (a^\dag_{1})^{n_1} (a^\dag_{2})^{n_2}
\ldots \ket{\Omega} .
\end{equation}
The same formalism can be adopted in the Ising case by setting:\footnote{In $1d$ this
transformation can be made more rigorous using the
Jordan-Wigner transformation.} $a^\dag_i \simeq
S^+_i $ and taking $\ket\Omega \equiv
\ket{\downarrow\ldots\downarrow}$. Here we can define an excitation with
defined momentum by setting:
$$ 
\aaaa^\dag_k = \sum_{j = 0}^{L-1} e^{i k j} a_j^\dag. 
$$
Putting the inverse of this expression
into Eq. (\ref{secondquantizationrealspace}), we get a new basis of eigenstates of
total momentum $P$:
\begin{equation}
 \label{secondquantizationfourier}
 P \aaaa^\dag_{k_1}\aaaa^\dag_{k_2}
\ldots \ket{\Omega} = \left(\sum_i k_i\right)
\aaaa^\dag_{k_1}\aaaa^\dag_{k_2}
\ldots \ket{\Omega}.
\end{equation}
We call this the \textit{single-particle Fourier basis} (SPFB).
However once we restrict to a subspace where $P$ is defined (appearing as a
block for the Hamiltonian matrix), there is a strong difference between this
case and the one defined in Eq. (\ref{secondquantizationrealspace}): in fact here,
not only the full state but each component excitation has a defined momentum. To
better understand this difference we consider a two particle case (in
the one particle case, there are no differences between the two momentum bases).  Consider
a state $\ket s = S^\dag_{x_1} S^\dag_{x_2} \ket{\Omega} = \ket{x_1,x_2}$, that
is the state with only two up spins in positions, $x_1$ and $x_2$. In the RTFB, this
state can be expressed as a linear combination of $L$ states, $\ket{\tilde s_n}$, with $n = 0,\ldots, L-1$,
where $|\tilde s_n\rangle$ are obtained as superpositions of the rigid translations, $T^k \ket{s} $.
In these translated states the 
distance $|x_2 - x_1|$ between up states always remains the same. On the other hand, with the SPFB, each
two-particle state has a non-zero matrix element with $\ket{x_1,x_2}$:
$$ \bra{x_1,x_2}\aaaa_{k_1}^\dag \aaaa_{k_2}^\dag
\ket\Omega = \sum_{j_1,j_2} e^{i (k_1 j_1 + k_2 j_2)} \bra{x_1,x_2}
j_1,j_2\rangle = e^{i (k_1 x_1 + k_1 x_2)} + e^{i (k_1 x_2 + k_1 x_1)}. $$
Increasing the number of particles to $M$, in the RTFB case we always have one
summation with $\simeq L$ terms. Instead in the SPFB, we will have $M$
summations, corresponding to $\simeq L^M$ terms.

The conclusions we can draw from these considerations are that:
\begin{itemize}
 \item A local Hamiltonian will appear as a sparse matrix in the real-space
basis.
 \item If we consider the Fourier basis obtained as a superposition of the rigid
translations of the real-space basis, the Hamiltonian will appear again as a
sparse matrix though perhaps with a larger density of non-zero entries. This is true
simply because the change of basis we are considering is sparse.
 \item If we consider the Fourier basis obtained taking the Fourier transform
of single particle states, the change of basis in each block of defined
total momentum will not be sparse.  In the general case, the
Hamiltonian matrix will be characterized by dense blocks of fixed total
momentum. 
\end{itemize}

\section{Generation of the mask}
\label{appendixMask}
Our aim is to generate a mask, telling us where the non-zero entries are, according to the scaling Eq. (\ref{densityones}).
The most obvious choice would be simply to take the mask such that each
non-diagonal entry is chosen independently, being one with probability $\rho =
\frac{\ln N}{N}$ and zero otherwise: $\rho_0 = 1 - \rho$. But a simple
computation shows that with this choice the probability $P_{nc}$ of having a
column (or row) with all the non-diagonal entries equal to zero will be quite
high
$$ 
P_{nc} \simeq 1 - \left(1 - \rho_0^N \right)^N \xrightarrow{N\to\infty}
\frac{e-1}{e} \simeq 63\% .
$$
To avoid this, we try to fix the number of non-zero entries in each row. However, since at the end, 
we will need a symmetric matrix with one on the diagonal, we follow a slightly more complicated procedure. 
We first generate ``half'' of the mask matrix and at the end we symmetrize by summing it with its transpose. 
To be more precise, for each row, $i$, of the matrix we generate a set, $R_i$, composed of integers
drawn from the set $\{1,\ldots,\hat i,\ldots,N\}$, i.e. a set where the element $i$ is missing.
The size, $n_i$, of $R_i$ has probability
$P_{in}$ of equaling $n$ where we define $P_{in}$ by
$$ 
P_{in} = \{q\}\delta_{n-1,\lfloor q\rfloor}  + (1 - \{q\})\delta_{n , \lfloor q \rfloor} 
$$
where $\{q\}$ and $\lfloor q \rfloor$ represent respectively the fractional and integer part of $q$ and $q$ is defined by:
$$ 
q = (N-1) - \sqrt{(N-1)^2 - (N-1) \ln N} .
$$
In this way, the size of the $R_i$'s will be either $\lfloor q \rfloor$ or $\lfloor q \rfloor +1$, but such that the average
is  $\bar n_i=q$. The sets $R_i$'s contain the indexes of the non-zero elements in each row. 
We now define the mask joining each row with the corresponding column, thus symmetrizing the matrix:
$$ \mathcal{M}_{ij} = \left\{\begin{array}{ll}
                            1 & (i = j)\vee (j \in R_i) \vee (i \in R_j);\\
                            0 & \text{otherwise}.
                           \end{array}\right. 
$$
Our choice of $q$ takes into account the possibility of elements overlapping between the row and the corresponding column, ensuring we have the correct number of non-zero elements in each row:
$$ 
\overline{ \# \{ j \; | \; \mathcal{M}_{ij} \neq 0 \}}  = 1 + 2 q -
\frac{q^2}{N-1} = 1 + \ln N.
$$
Notice that the number of non-zero elements in each row is not fixed, because 
small fluctuations are induced by the symmetrization procedure; however we can be sure that it will be greater than $\lfloor q \rfloor$.
We recall that $\ln N / N$ is exactly the threshold for the connectedness of 
the Erdos-Renyi graph \cite{Erdos}: with this procedure there will be no
isolated subblocks, as is confirmed by the Wigner-Dyson level statistics shown in Fig. \ref{dosspacingsparse}.

\end{document}